\documentclass[journal]{IEEEtran}
\ifCLASSINFOpdf
\else
\fi

\usepackage{cite}
\usepackage{amsmath,amssymb,amsfonts}
\usepackage{graphicx}
\usepackage{color}
\usepackage{stackengine}
\usepackage{comment}
\usepackage[normalem]{ulem}
\usepackage{tabularx}
\usepackage{bm}
\usepackage{dsfont}
\usepackage{mathtools}
\usepackage[font=scriptsize]{subcaption}
\usepackage{lettrine}
\usepackage[font=scriptsize]{caption}

\usepackage[flushleft]{threeparttable} % http://ctan.org/pkg/threeparttable
\usepackage{booktabs}

\usepackage{textcomp}
\usepackage{xcolor}

\usepackage[linesnumbered, ruled]{algorithm2e}
\SetKwRepeat{Do}{do}{while}%

\SetCommentSty{mycommfont}
\usepackage{algpseudocode}

\def\BibTeX{{\rm B\kern-.05em{\sc i\kern-.025em b}\kern-.08em
    T\kern-.1667em\lower.7ex\hbox{E}\kern-.125emX}}
    
\usepackage{orcidlink}

\usepackage{xspace}
\usepackage{ifthen}
\usepackage[inline]{enumitem}

\usepackage{algcompatible}

\usepackage{pifont}     % For \ding
\usepackage{xcolor}     % For colored text like \textcolor
\usepackage{placeins}

\newcommand{\cmark}{\textcolor{green}{\ding{51}}} % Green check mark
\newcommand{\xmark}{\textcolor{red}{\ding{55}}}   % Red cross mark

\newboolean{showcomments}
\setboolean{showcomments}{true}
\ifthenelse{\boolean{showcomments}}
{
}

\usepackage{fancyhdr}

\begin{document}
%\title{\huge Leveraging Instance Segmentation and Multimodal-LLM for Traffic Monitoring}
%\title{\huge Efficient Traffic Surveillance Using Vision Transformers and LLMs: An Edge–Cloud Semantic Communication Approach  }
\title{Semantic Edge–Cloud Communication for Real-Time Urban Traffic Surveillance with ViT and LLMs over Mobile Networks}
%\title{ViT-GEN-SAM: ViT-based Generative AI for Semantic traffic surveillance over Mobile networks}
\author{
\IEEEauthorblockN{Murat Arda Onsu$^1$, Poonam Lohan$^1$, Burak Kantarci$^1$, Aisha Syed$^2$, Matthew Andrews$^2$, Sean Kennedy$^2$}\\
\IEEEauthorblockA{\textit{$^1$University of Ottawa, Ottawa, ON, Canada}\\
\textit{$^2$Nokia Bell Labs, 600 March Road,
Kanata, ON K2K 2E6, Canada}\\
$^1$\{monsu022, ppoonam, burak.kantarci\}@uottawa.ca,~$^2$\{aisha.syed, matthew.andrews, sean.kennedy\}@nokia-bell-labs.com}
}

\maketitle
\begin{abstract}
Real-time urban traffic surveillance is vital for Intelligent Transportation Systems (ITS) to ensure road safety, optimize traffic flow, track vehicle trajectories, and prevent collisions in smart cities. Deploying edge cameras across urban environments is a standard practice for monitoring road conditions. However, integrating these with intelligent models requires a robust understanding of dynamic traffic scenarios and a responsive interface for user interaction. Although multimodal Large Language Models (LLMs) can interpret traffic images and generate informative responses, their deployment on edge devices is infeasible due to high computational demands. Therefore, LLM inference must occur on the cloud, necessitating visual data transmission from edge to cloud, a process hindered by limited bandwidth, leading to potential delays that compromise real-time performance. To address this challenge, we propose a semantic communication framework that significantly reduces transmission overhead. Our method involves detecting Regions of Interest (RoIs) using YOLOv11, cropping relevant image segments, and converting them into compact embedding vectors using a Vision Transformer (ViT). These embeddings are then transmitted to the cloud, where an image decoder reconstructs the cropped images. The reconstructed images are processed by a multimodal LLM to generate traffic condition descriptions. This approach achieves a 99.9\% reduction in data transmission size while maintaining an LLM response accuracy of 89\% for reconstructed cropped images, compared to 93\% accuracy with original cropped images. Our results demonstrate the efficiency and practicality of ViT and LLM-assisted edge–cloud semantic communication for real-time traffic surveillance.

\end{abstract}


\section{Introduction} \label{sec:1}

Intelligent and automated urban traffic analysis has been extensively studied for various applications, including traffic rule violation detection, collision prevention, road safety enhancement, vehicle tracking, and congestion management through route suggestions. Real-time camera feeds serve as a critical source of data, capturing key traffic information such as vehicle count, positions, trajectories, congestion levels, and potential collisions. This data can be processed using intelligent models to generate appropriate responses to traffic-related queries, thereby ensuring both safety and operational efficiency \cite{sur1, sur2}. To be truly effective in real-time urban environments, these models must possess a deep understanding of dynamic traffic conditions and deliver fast, accurate responses through an intuitive and user-friendly interface.

Multimodal large language models (LLMs) offer a powerful solution for urban traffic monitoring by enabling the interpretation of both visual data and textual queries to generate meaningful, human-readable responses \cite{wu2023multimodal, nr5,Friha.2024}. Unlike traditional text-only LLMs, these models can effectively process camera images alongside user queries, making them highly suitable for complex traffic analysis tasks. However, the substantial computational and memory requirements of multimodal LLMs make them impractical for deployment on resource-constrained edge devices, such as surveillance cameras. To address this limitation, our framework leverages edge-cloud architecture \cite{Ferrag.2023} and images captured at the edge are transmitted to the cloud, where a multimodal LLM is deployed for efficient traffic monitoring and response generation.
In edge-cloud architecture, distributed edge devices such as surveillance cameras, which reside close to the data collection points for low-latency objectives, integrate with the centralized cloud servers having high-performance capabilities. The real-time collected data from the edge devices performs the required lightweight task and communicates with the centralized cloud, which provides high resources and scalability for a given task.% An edge platform also supports temporal and spatial synchronization of data across distributed nodes, ensuring coherent interpretation of multi-source inputs. }

However, transmitting high-resolution camera images to the cloud in real-time can lead to memory overload and latency issues due to transmission bottlenecks, limited bandwidth, and constrained network capacity at the edge \cite{sur3}. To address this challenge, semantic communication has emerged as a promising solution for reducing bandwidth consumption during data transmission. Unlike Shannon’s classical communication theory, which emphasizes the accurate transmission of bits and symbols, semantic communication focuses on conveying only the information that is meaningful and relevant to the receiver’s specific task \cite{17}. Recent research highlights the effectiveness of semantic communication in Intelligent Transportation System (ITS) applications, including the Internet of Vehicles (IoV), traffic perception, and real-time traffic management \cite{3,5}. Furthermore, by leveraging the attention mechanisms in transformer architectures \cite{4}, semantic communication systems can dynamically identify and prioritize the most informative parts of the input data. This enables the selective transmission of task-relevant content, significantly reducing communication overhead while preserving the semantic integrity of the information \cite{19}.

In this research, we build upon our previous work \cite{onsu2025leveraging} on traffic monitoring by incorporating a semantic communication framework that integrates both the Vision Transformer (ViT) \cite{22, more_1_vit, more_2_vit} and the object detection model YOLO (You Only Look Once) \cite{23, more_1_yolo, more_2_yolo}. The proposed approach begins with a pre-trained YOLO model applied to incoming camera images to identify and highlight Regions of Interest (RoIs). To reduce data redundancy and facilitate more efficient processing by multimodal LLM, only these critical regions are cropped and used for downstream tasks. At the edge, the cropped image segments are encoded into embedding vectors using ViT and then transmitted to the cloud. On the cloud server, these embedding vectors are reconstructed into images using a deep learning (DL) decoder built with Transposed Convolutional Neural Network (Transposed-CNN) layers \cite{20}. The reconstructed images are then processed by a fine-tuned multimodal LLM, LLaVA 1.5 7B \cite{24, more_1_llava, more_2_llava}, for generating traffic-related responses. The primary contributions of this work are summarized as follows:

\begin{itemize}
    \item This study presents a comprehensive end-to-end pipeline that integrates YOLOv11 for vehicle detection, ViT for embedding generation, a custom image decoder for image reconstruction, and LLaVA for language-based traffic description. The proposed system enhances both communication efficiency and AI-driven interpretability for smart city surveillance applications.
    \item A novel semantic communication framework is introduced, which significantly reduces transmission data size by approximately 99.9\% by transmitting ViT-generated embedding vectors of vehicle-centric RoIs instead of raw or cropped images from edge devices to the cloud.
    \item The Mean Square Error (MSE) of the reconstructed embedding vectors and the LPIPS score of the reconstructed cropped images are evaluated for both IEEE 754 floating-point and quantization-based bitstream encoding methods. The results demonstrate that the quantization technique offers greater robustness to bit errors while significantly reducing transmission data size.
    \item A lightweight multimodal large language model, LLaVA 1.5 (7B), is fine-tuned to support real-time urban traffic monitoring by generating accurate and timely responses based on vehicle-centric cropped images. The model attains 89\% accuracy when processing reconstructed images from embedding vectors, in comparison to 93\% accuracy achieved with the original cropped images.  
\end{itemize} 

The rest of the paper is organized as follows. Section \ref{sec:2} presents related works. Section \ref{sec:methodology} describes the methodology, including dataset creation, embedding vector generation, image reconstruction, and multimodal-LLM responses. Section \ref{sec:results} explains the training process and numerical results for memory savings and multimodal-LLM performance with visual examples, and Section \ref{sec:conclusion} concludes the paper.

\section{Related Works} \label{sec:2}
Recent advancements in semantic image communication introduce novel frameworks and evaluation metrics that significantly enhance the efficiency and semantic fidelity of image transmission over wireless channels. These efforts are particularly relevant for bandwidth-constrained environments such as intelligent traffic systems. During literature review, our main criteria are the efficient way for image transmissions, compressions, reconstruction quality, via several methods under different SNR ratios.

To optimize resource utilization and ensure precise transmission of critical information, the importance of image patches is assessed by combining objective and subjective user attention, and the proposed model is tested under various channel conditions. The study in \cite{nr15} proposes a semantic V2X (Vehicle to Everything) communication system for image transmission in 6G networks, aimed at addressing the increasing demands of vehicular communication scenarios. It utilizes a semantic encoder-decoder architecture and introduces a semantic importance evaluation mechanism to selectively transmit critical visual content. The work in \cite{nr16} presents a multiuser cooperative semantic communication framework for the Internet of Vehicles (IoV). It focuses on jointly optimizing semantic extraction and resource allocation to improve transmission efficiency and enables vehicles to share semantic features through cooperative transmission, to reduce latency, and achieve higher robustness in multi-user vehicular networks.  

\subsection{Deep Learning-Assisted Semantic Communication}

In \cite{1}, the DeepSC-RI framework enables robust image (RI) transmission by exploiting multi-scale semantic information to mitigate semantic impairments and improve fidelity. Similarly, the ISSC (Image Segmentation Semantic Communication)  system in \cite{12} targets IoV scenarios, using a Swin Transformer-based multi-scale feature extractor to encode semantic features of road images and reconstruct segmentation maps. This approach significantly reduces transmission overhead while maintaining high mIoU and robustness against noise. In \cite{mr12}, a personalized saliency-driven framework is introduced for semantic image transmission, where only the most relevant regions are encoded and sent, significantly lowering bitrate without losing essential content. To further reduce bandwidth, \cite{mr7} propose an image transmission system that stores static backgrounds at the receiver and transmits only regions of interest extracted via YOLO. These regions are encoded using 36 bits for bounding box coordinates and 8 bits per pixel for cropped image data, with Polar Codes and BPSK modulation over an AWGN channel. Normalized MSE and SSIM are used as metrics, showing an 86.72\% bandwidth saving. Moreover, Researchers in \cite{mr5} propose DeepMA, a deep learning–based multiple access method for semantic communication, designed to reduce complexity and improve compatibility. Their model is evaluated on wireless image transmission tasks using the CIFAR and Kodak datasets under SNR levels between 0–20 dB. In \cite{mr16}, semantic communication is applied to vehicle count prediction in smart cities, enabling central traffic controllers to make efficient real-time decisions. The approach transmits only task-relevant features such as vehicle density maps, using a CNN–LSTM encoder–decoder for enhanced feature representation, and is evaluated with MSE and MAE for vehicle count accuracy and density estimation.

For connected autonomous vehicles, \cite{mr2} present a multitask semantic communication framework robust to adverse weather conditions such as snowstorms. Their model applies convolutional autoencoder–based semantic encoding to traffic sign images observed by leading vehicles, which are then shared among other vehicles. The framework supports both task-oriented image reconstruction and classification, with performance tested across varying SNR levels. In \cite{nr1}, researchers focus on semantic impairments in images to address the transmission bottlenecks and propose a metric to quantify the intensity of semantic impairment. Moreover, the introduction of DL enables a semantic communication system, DeepSCRI, to leverage multi-scale semantic information to mitigate semantic impairments substantially. Moreover, to enable high-level conceptual semantic embedding, the authors in \cite{nr4} propose a Domain Knowledge-driven Semantic Communication (DKSC) framework. This dual-path architecture incorporates semantic extraction and reconstruction at both the information and conceptual levels. Leveraging deep neural network-based encoders and decoders, the DKSC system supports end-to-end wireless communication and image transmission over wireless channels. A semantic base-enabled image transmission framework that incorporates fine-grained Hybrid Automatic Repeat Request (HARQ) mechanisms is presented in \cite{nr11} to enhance image fidelity and robustness while minimizing communication resource consumption in wireless environments. The proposed system analyzes image semantics to prioritize and transmit the most relevant features, while the HARQ strategy selectively retransmits semantically important components under poor channel conditions. Another semantic communication system for image transmission is proposed in \cite{nr12} that can distinguish ROI and non-ROI regions based on semantic segmentation and enables high-quality transmission of ROI with lower communication overheads using a proper data compression algorithm.

The paper \cite{nr17} introduces ITS-SC, a semantic communication model tailored for vehicle-infrastructure collaboration systems (i-VICS). It employs Deep JSCC and integrates the Swin Transformer architecture to extract and transmit image semantics efficiently. This design improves image reconstruction quality under low bandwidth and noisy conditions. Scene graph-guided semantic transmission, proposed in \cite{14}, structures semantic information using scene graphs, allowing for context-aware and relationship-driven encoding beyond raw pixel representation. In \cite{16}, a deep JSCC (Joint Source-Channel Coding) framework jointly optimizes image recovery and classification using a unified loss function that combines MSE (Mean Square Error) minimization with coding rate reduction. A gated JSCC approach with domain randomization further enhances robustness under varying channel conditions. Reinforcement learning is explored in \cite{18} through the RL-ASC (reinforcement learning-based adaptive semantic coding) framework, which encodes semantic concepts such as category, spatial arrangement, and visual features using a convolutional encoder guided by a reward function that balances transmission rate, semantic relevance, and perceptual quality.

\subsection{Generative Model-Assisted Semantic Communication}

To improve the trade-off between visual perception and compression, a coarse-to-fine image compression framework in \cite{6} leverages a GAN-based generator to reconstruct a perceptually rich base layer while enhancing fine details through BPG residual coding. The reconstruction quality is quantitatively evaluated using a combination of perceptual and semantic metrics, including FID (Fréchet Inception Distance), KID (Kernel Inception Distance), and LPIPS (Learned Perceptual Image Patch Similarity) for visual fidelity, along with mIoU (mean Intersection-over-Union) to assess semantic consistency. Researchers in \cite{mr4} propose a deep conditional GAN–based semantic communication system for image transmission at low SNR to overcome the cliff effect, employing JSCC to balance distortion, perception, and rate. Their model combines pixel-level MSE with VGG-based perceptual loss and achieves improved visual realism, particularly in low-SNR conditions. In \cite{mr8}, a semantic communication framework is developed for next-generation networks such as ITS, where intelligent semantic extraction, mobile segment analysis, and GAN-based reconstruction with denoising are employed to reduce redundant transmissions. The framework uses adversarial and L1 losses and demonstrates robustness under SNR levels from 0–10 dB. To enhance semantic accuracy for unmanned surface vehicles, \cite{mr11} propose the SC-IRT algorithm, which integrates an autoencoder-based encoder with a GAN-based decoder and multiscale discriminators. Their experiments on the WSODD dataset, evaluated with MSE, LPIPS, SSIM, and compression ratio, confirm improved reconstruction and efficiency. \cite{mr10} present a bandwidth-efficient and privacy-preserving framework using Semantic StyleGAN and RealNVP-based Normalizing Flow, where reconstruction quality is assessed with LPIPS. A GAN-based decoder ensures semantic consistency even under low bitrates and noisy conditions. A semantic communication-based end-to-end image transmission system is implemented in \cite{nr14}, and to reconstruct a realistic image from the semantic segmented image at the receiver input, a pre-trained GAN network is also utilized for transmission task. \cite{mr9} proposes a generative AI framework for handling large-scale multimodal vehicular data from cameras, LiDAR, and textual sensors, addressing transmission instability caused by dynamic networks. Experiments are conducted on the nuScenes-mini dataset under SNR levels ranging from –8 to 16 dB.

\begin{table*} []
\centering
\begin{tabular}{lccccc}
\multicolumn{1}{c}{{\color[HTML]{000000} \textbf{Method}}}                                                                                             & {\color[HTML]{000000} \textbf{LLM}}    & {\color[HTML]{000000} \textbf{\begin{tabular}[c]{@{}c@{}}Multi-\\ Modality\end{tabular}}} & {\color[HTML]{000000} \textbf{ViT}}    & {\color[HTML]{000000} \textbf{Segmentation}} & {\color[HTML]{000000} \textbf{\begin{tabular}[c]{@{}c@{}}Image\\      Reconstruction\end{tabular}}} \\ \hline
{\color[HTML]{000000} \begin{tabular}[c]{@{}l@{}}Federated   Learning Based \\ Semantic Communication (FLSC) \cite{5}\end{tabular}}                    & {\color[HTML]{000000} \xmark}          & {\color[HTML]{000000} \xmark}                                                             & {\color[HTML]{000000} \textit{\cmark}} & {\color[HTML]{000000} \xmark}                & {\color[HTML]{000000} \textit{\cmark}}                                                              \\ \hline
{\color[HTML]{000000} \begin{tabular}[c]{@{}l@{}}Vehicular Image Segmentation-Oriented \\ Semantic Communication (VIS-SemCom) \cite{12}\end{tabular}}  & {\color[HTML]{000000} \xmark}          & {\color[HTML]{000000} \xmark}                                                             & {\color[HTML]{000000} \textit{\cmark}} & {\color[HTML]{000000} \textit{\cmark}}       & {\color[HTML]{000000} \xmark}                                                                       \\ \hline
{\color[HTML]{000000} \begin{tabular}[c]{@{}l@{}}Context-Aware SemCom Framework \\ (CaSemCom) \cite{nr3}\end{tabular}}                                 & {\color[HTML]{000000} \textit{\cmark}} & {\color[HTML]{000000} \textit{\cmark}}                                                    & {\color[HTML]{000000} \xmark}          & {\color[HTML]{000000} \xmark}                & {\color[HTML]{000000} \xmark}                                                                       \\ \hline
{\color[HTML]{000000} \begin{tabular}[c]{@{}l@{}}Generative SemCom For Multi-User \\ System (M-GSC)   \cite{nr10}\end{tabular}}                        & {\color[HTML]{000000} \textit{\cmark}} & {\color[HTML]{000000} \xmark}                                                             & {\color[HTML]{000000} \xmark}          & {\color[HTML]{000000} \textit{\cmark}}       & {\color[HTML]{000000} \xmark}                                                                       \\ \hline
{\color[HTML]{000000} \begin{tabular}[c]{@{}l@{}}Reinforcement Learning Based Adaptive \\ Semantic Coding (RL-ASC) \cite{18}\end{tabular}}             & {\color[HTML]{000000} \xmark}          & {\color[HTML]{000000} \xmark}                                                             & {\color[HTML]{000000} \xmark}          & {\color[HTML]{000000} \xmark}                & {\color[HTML]{000000} \textit{\cmark}}                                                              \\ \hline
{\color[HTML]{000000} \begin{tabular}[c]{@{}l@{}}Task-Oriented Semantic Communication \\ Framework Using LLaVA   \cite{nr13}\end{tabular}}             & {\color[HTML]{000000} \textit{\cmark}} & {\color[HTML]{000000} \textit{\cmark}}                                                    & {\color[HTML]{000000} \textit{\cmark}} & {\color[HTML]{000000} \textit{\cmark}}       & {\color[HTML]{000000} \xmark}                                                                       \\ \hline
{\color[HTML]{000000} \begin{tabular}[c]{@{}l@{}}SemCom   in Intelligent Transportation \\ Systems \cite{nr15}\end{tabular}}                           & {\color[HTML]{000000} \xmark}          & {\color[HTML]{000000} \xmark}                                                             & {\color[HTML]{000000} \xmark}          & {\color[HTML]{000000} \textit{\cmark}}       & {\color[HTML]{000000} \xmark}                                                                       \\ \hline
{\color[HTML]{000000} \begin{tabular}[c]{@{}l@{}}Image Transmission System Based On \\ Semantic Communication,  (ITS-SC) \cite{nr17}\end{tabular}}     & {\color[HTML]{000000} \xmark}          & {\color[HTML]{000000} \xmark}                                                             & {\color[HTML]{000000} \textit{\cmark}} & {\color[HTML]{000000} \xmark}                & {\color[HTML]{000000} \textit{\cmark}}                                                              \\ \hline
{\color[HTML]{FF0000} \begin{tabular}[c]{@{}l@{}}Semantic Edge–Cloud Communication with \\ Segmentation Assisted ViT and LLMs (Our Work)\end{tabular}} & {\color[HTML]{000000} \textit{\cmark}} & {\color[HTML]{000000} \textit{\cmark}}                                                    & {\color[HTML]{000000} \textit{\cmark}} & {\color[HTML]{000000} \textit{\cmark}}       & {\color[HTML]{000000} \textit{\cmark}}                                                              \\ \hline
\end{tabular}

\vspace{0.15cm}
\caption{Qualitative comparison of the related works in the literature and this paper}
\label{tab:tab_literature}

\end{table*}

\subsection{ViT-Assisted Semantic Communication}

Vision Transformers (ViTs) emerge as powerful tools in semantic communication. The ViT-based model in \cite{13} achieves a +0.5 dB PSNR gain over CNN-based variants and introduces metrics such as average cosine similarity and Fourier analysis to analyze internal model dynamics. In \cite{19}, attention masks guide ViTs to prioritize semantically important regions of input images, improving compression efficiency and reconstruction quality under bandwidth constraints. A custom ViT architecture in \cite{21} demonstrates superior performance and robustness compared to CNNs and GANs across datasets like ImageNet and CIFAR, achieving up to 72\% bandwidth savings under challenging channel conditions. To support distributed and scalable semantic transmission, a federated learning-based semantic communication (FLSC) framework is proposed in \cite{5} for multitask image transmission across IoT devices. It uses a hierarchical vision transformer (HVT)-based extractor and a task-adaptive translator for coarse-to-fine semantic representation. Channel state information-based MIMO modules further help address fading and noise effects. 

In \cite{mr1}, the objective is to improve image transmission in complex IIoT channels. The method employs end-to-end semantic communication using a ViT for feature extraction, reducing bandwidth consumption. Results show improved reconstruction quality, measured with PSNR and MS-SSIM under varying channel conditions. In \cite{mr6}, a channel-aware adaptive framework is proposed for image encoding and transmission. The model combines ViT with a CNN-based encoder–decoder for patch mapping. Performance is evaluated by MSE and classification accuracy across different resolutions. In \cite{mr3}, to enhance robustness and mitigate semantic impairment they proposed DeepSC-RI model, uses a multi-scale ViT with coarse and fine grained branches for semantic extraction. Evaluation metrics include LPIPS, PSNR, and classification accuracy, showing improved resilience to semantic loss. ViTScore in \cite{10} serves as a semantic similarity metric based on pre-trained Vision Transformers. It leverages attention mechanisms to quantify semantic consistency while satisfying key properties like symmetry and normalization. ViTScore outperforms traditional metrics across tasks involving semantic communication and adversarial attacks. As shown in the qualitative comparison, the proposed work in this paper stands out by utilizing multimodal-LLM with ViT assisted by a segmentation model into the overall solution.

\subsection{LLM-Assisted Communication}

The multimodal-LLM model is integrated in \cite{nr2} alongside advanced DL methods for comprehensive road perception in urban traffic environments. They first deploy object detection models for traffic sign recognition. A lightweight multimodal-LLM framework is designed to enhance adaptive lane detection through contextual reasoning capabilities in autonomous driving. Context-Aware Semantic Communication, CaSemCom, is proposed in \cite{nr3} to overcome the overlooking of dynamic conditions and contextual cues. By leveraging LLM-based gating mechanisms and a Mixture of Experts (MoE), CaSemCom adaptively selects and encodes only high-impact semantic features across multiple data modalities. In \cite{nr10}, researchers explore a generative semantic communication framework for multi-user systems, M-GSC, with LLMs as the shared knowledge base to enhance semantic extraction and reasoning. The proposed approach targets improving performance in shared wireless environments by leveraging LLMs' generative capabilities to capture context and intent more effectively than traditional methods. 

Research \cite{nr9} proposes an LLM-driven semantic communication framework for edge-based IoT networks to enhance communication efficiency by extracting and transmitting task-relevant information. Their proposed framework uses LLM for semantic communication in edge based IoT system to enhance the performance and utility of LLMs by bringing computation and data storage closer to the data sources. Researchers, in \cite{nr7}, propose LAM-MSC, a large AI model-based framework for multimodal semantic communication to address data heterogeneity, semantic ambiguity, and signal distortion by leveraging multimodal-LLM. In their framework, LLMs are used for personalized knowledge extraction, while multimodal language models are used for modality alignment. Moreover, they apply conditional generative adversarial networks-based channel estimation to estimate the wireless channel state information. In \cite{nr8}, an OFDM-based semantic communication framework is proposed for image transmission using LLMs. The system introduces a semantic encoder that leverages LLMs to extract and transmit only the essential meaning of images, rather than raw data. On the receiver side, an LLM-based decoder reconstructs the image contextually. Researchers in \cite{k2} propose a task-oriented semantic communication framework for vehicle networks leveraging LMMs such as LLaVA. Their method integrates YOLO for object detection and GloVe for matching user query keywords with detected labels, aiming to reduce computational load, prioritize important features, and optimize resource allocation. Performance is evaluated using bit error rate and task accuracy. Researchers in \cite{nr6} propose LaMoSC, a semantic communication system that integrates Large Language Models (LLMs) with visual-textual multimodal fusion to improve image transmission quality under low signal-to-noise ratio (SNR) conditions. Their model features an end-to-end encoder-decoder architecture combining ViT and LLM-based prompt text extraction to fuse textual and visual features, enhancing robustness and generalization.

Researchers in \cite{k1} propose the LLM-DiSAC framework to overcome the limitations of single-model systems, such as limited spatial coverage, high communication cost, and weak semantic understanding. Their design integrates a Hybrid CNN–ViT vision extractor, semantic encoder with LLaMA-3 decoder, and a Transformer-based cross-device aggregator, evaluated by NMSE under AWGN across 0–25 dB SNR. In \cite{k3}, an LLM-supported semantic communication system for image transmission is studied, where LLaVA serves as the semantic encoder and CLIP as the decoder to reduce data size while preserving contextual meaning over wireless channels with QAM modulation; performance is measured using compression ratio and image quality. To improve bitrate efficiency and semantic robustness under noisy conditions, \cite{k5} combine LLM reasoning, generative diffusion models, and adaptive offloading strategies. A multimodal LLM-integrated framework is introduced in \cite{k6} to lower transmission costs by extracting task-relevant features from multimodal inputs. Their system employs LLaVA, Vision Transformers, and CLIP on the encoder side, and a resource-adaptive decoder built with a Variational Autoencoder and conditional diffusion model for image reconstruction. Two case studies, VQA and image generation, are evaluated using PSNR and LPIPS under SNR levels of 5, 7, 9, and 12 dB. Research in \cite{nr13} introduces a task-oriented semantic communication framework for vehicle networks built upon the LLaVA model to facilitate efficient interaction between users and cloud servers.  Relevant literature on semantic communication for urban traffic scenarios is summarized in Table \ref{tab:tab_literature}.

These collective efforts lay a strong foundation for integrating ViTs and LLMs in semantic communication systems for traffic surveillance. Our work builds upon these innovations by proposing a hybrid edge–cloud semantic communication framework that utilizes ViTs for semantic encoding and LLMs for contextual reasoning in traffic scene understanding. Moreover, this method is assisted by an object detection model, YOLO, in collecting RoI parts from the image before ViT to reduce the transmission data of the semantic communication.


\section{Methodology} \label{sec:methodology}

\begin{figure*}
    \centering
    \includegraphics[width=0.8\linewidth]{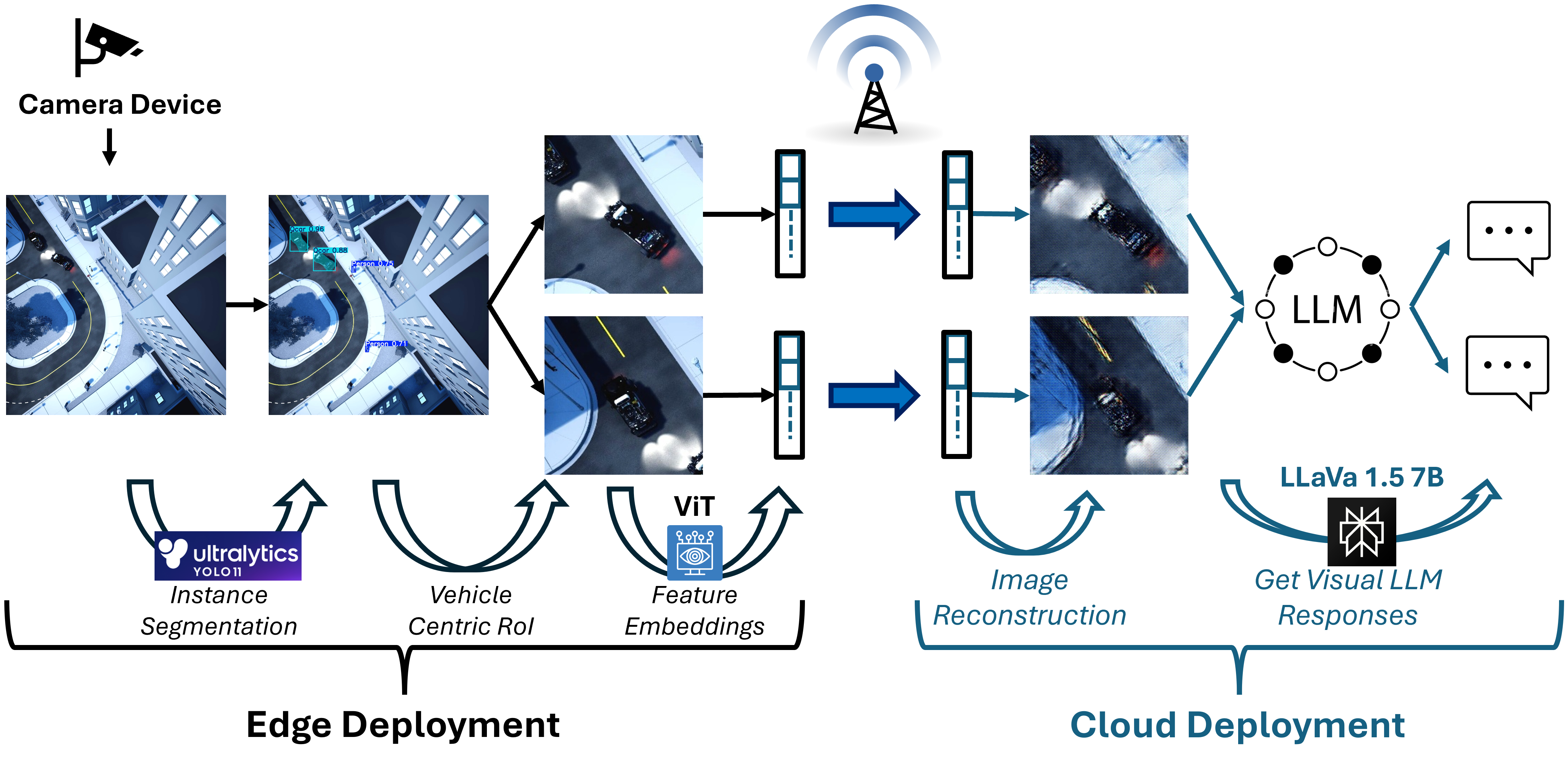}
    \caption{Overall Workflow Semantic Communication and Traffic Monitoring Pipeline, Including Instance Segmentation, Vision Transformer and LLaVA Model}
    \label{fig:workflow}
\end{figure*}

In this research, the objective is to perform intelligent traffic monitoring using edge camera captions on the multimodal LLM, which resides on the cloud server.  During the image transmission from the edge device to the cloud, the target is to leverage the semantic communications to reduce the bandwidth consumption and save memory. The initial step is to utilize the YOLO model to get the segment coordinates of each vehicle. Then, each vehicle,  detected by the YOLO model, and its small windows of surroundings are cropped as an ROI area and forwarded to the ViT models to receive their embedded vector representations. These embedding representations are transmitted over a wireless channel to the cloud, where they are utilized to reconstruct the cropped images using an image decoder model comprising a Transposed Convolutional Neural Network (T-CNN) layer. The reconstructed images are then forwarded to the multimodal LLM, LLaVA, which generates responses to the corresponding traffic observation queries. The complete proposed workflow is depicted in \figurename \ref{fig:workflow}.   

Note that context-aware semantic communication frameworks \cite{nr3} have demonstrated strong performance in scenarios where environmental or application-specific context can be accurately modeled and exploited. However, this study focuses on a vision-based edge–cloud setting in which context information beyond the image content is either unavailable or highly dynamic. In our case, the data originates from a simulation platform capturing multi-vehicle traffic scenes without persistent metadata such as driver intent, network conditions, or geographic context, which are typically leveraged in context-aware systems. Furthermore, incorporating a context dimension would require an additional context acquisition and synchronization mechanism between the edge and cloud, potentially introducing latency and complexity that counteract the low-latency objective of this work. As the main aim of this work is to evaluate the efficiency of transmitting compact visual embeddings extracted from cropped ROIs, we limit the scope to content-based semantic communication.

\subsection{Data Collection using Quanser Interactive Lab} \label{dataset}

Edge traffic surveillance camera images and their textual captions create the structure of the database for our tasks. To provide flexibility and high-quality images, the Quanser Interactive Lab (Qlab) \footnote{Quanser, *Quanser Python API*, version 2024.10.17, 2024. Available at: \url{https://docs.quanser.com/quarc/documentation/python/getting_started.html}} simulation platform is utilized, which can be regarded as a digital twin of an autonomous vehicle track prototype. In the context of ITS, a digital twin refers to a high-fidelity virtual representation of a physical asset or environment that is continuously synchronized with real-world data. Such models enable safe, repeatable, and cost-effective testing of control strategies, perception algorithms, and traffic management policies before deployment in real-world conditions ~\cite{gu2025digital, irfan2024toward, hossain2023new, ge2024digital, wang2022automatic, zapata2025efficient}. In this framework, the Qlab replicates urban city environments with various components and actors, such as pedestrians, cars, sidewalks, surveillance cameras, traffic lights, and signs, which are programmed using either Python or MATLAB. These components can be spawned and placed anywhere in the simulation, allowing us to diversify the scenario and overcome real-world challenges during setup, such as changing the locations of traffic lights and signs, enrolling vehicles and pedestrians, and performing and repeating collision events. This facilitates the evaluation of computer vision pipelines and cloud-based inference systems for traffic management applications, consistent with the digital twin paradigm in ITS research.

%which provides urban city environments with various components and actors such as pedestrians, cars, sidewalks, surveillance cameras, traffic lights, and signs, programmed by either Python or MATLAB. These components can be spawned and placed anywhere in the simulation, allowing us to diversify the scenario and overcome real-world challenges during setup, such as changing the locations of traffic lights and signs, enrolling vehicles and pedestrians, and performing and repeating collision events repeatedly. Simulation environment and car actors are shown in \figurename \ref{fig:environment}.

\begin{figure}
    \centering
    \includegraphics[width=0.85\linewidth]{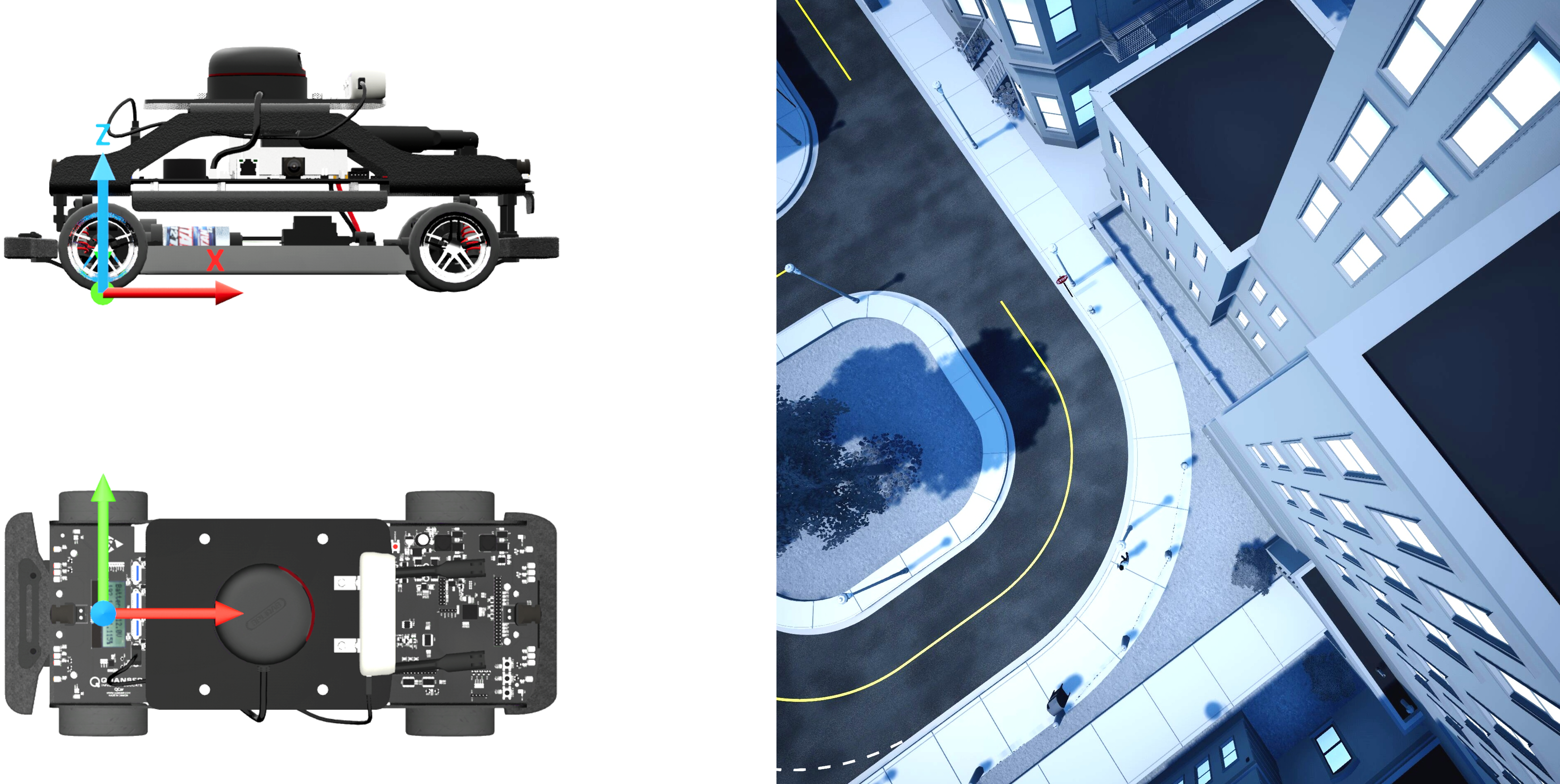}\vspace{0.2cm}
    \caption{Quanser Interactive Lab and QCar}
    \label{fig:environment}
\end{figure}

The right half of \figurename \ref{fig:environment} shows the empty traffic environments. At the same time, the left part illustrates the Quanser Simulation Car (Qcar), which is the leading actor for all scenarios. Before the simulation run, each Qcar's initial locations and routes are predetermined, and surveillance edge cameras are attached at various places in the environment, which comprise the overall area and never overlap to prevent duplication. In each simulation startup, the system resets and cameras collect the data, which are 2048x2048x3 RGB images. Each captured image is marked with a timestamp, the time at which they are collected, to ensure temporal alignment across cameras and enable accurate tracking of dynamic events \footnote{Urban Mobility Research Dataset (Generated with Quanser Interactive Lab), Available at: \url{https://dx.doi.org/10.21227/t4xs-rf05}}. After collecting camera images, this data is sent to the object detection and instance segmentation module to locate each Qcar location on the image and continue with further steps.

\subsection{Vehicle Captioning Using YOLOv11}

\begin{figure}
    \centering
    \includegraphics[width=0.85\linewidth]{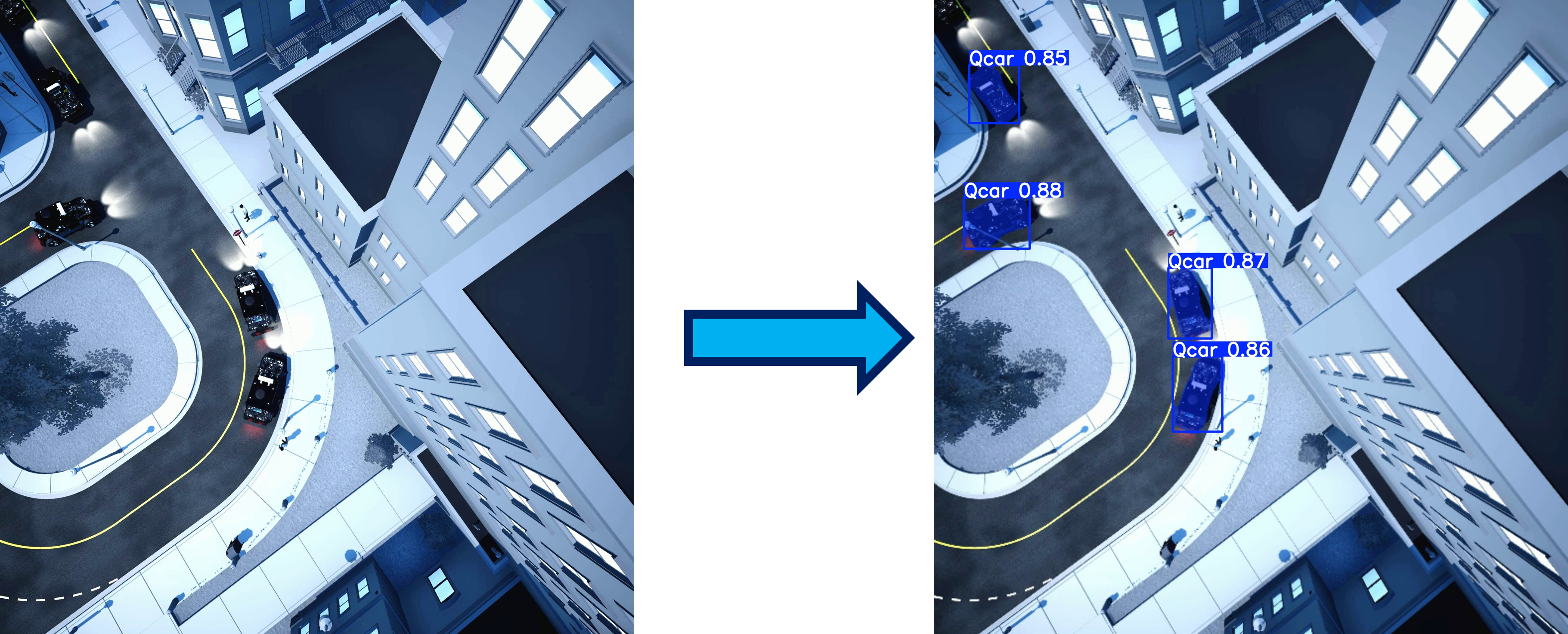}\vspace{0.2cm}
    \caption{YOLOv11 Model Output Sample}
    \label{fig:yolo_run}
\end{figure}

After road images are collected from edge cameras, these images are sent to the object detection and image segmentation module called the YOLOv model, which is trained to capture the Qcar actors on the simulation images. It is essential to locate these semantic features, which include the vehicles, and omit the redundant parts such as buildings, sky, and parks from the camera images to ease the multimodal-LLMs' work and provide the most necessary outputs. By employing the YOLO module, vehicles in the images are efficiently detected and tracked, allowing the LLM to concentrate on the most relevant regions. Built upon CNN layers, YOLO offers rapid processing speed and accurately identifies objects across the entire scene in a single pass. An example of its object detection and segmentation output is shown in \figurename \ref{fig:yolo_run}.

YOLOv11 version, which is an extension version of YOLOv8, is selected for this research, which is an evolution of the YOLO algorithm that represents a significant improvement in the real-time object detection task \cite{23}. This YOLO model architecture is formed with three basic components: Backbone, neck, and head. In the Backbone part, raw image data is processed by the CNN layer and transformed into multi-scaled feature maps for feature extraction. On the other hand, the neck component, which acts as an intermediate step, employs special layers to aggregate and enhance feature representation across different scales. Finally, prediction parts that generate proper outputs for object detection and localization are established in the head components.

YOLOv11 model is also used for instance segmentation task as well as object detection. 
The model integrates essential architectural elements, including the Spatial Pyramid Pooling - Fast (SPPF) module for multi-scale feature extraction and the Cross Stage Partial with Spatial Attention (C2PSA) module to improve spatial attention. These improvements enable YOLOv11 to capture fine details efficiently while maintaining computational efficiency \cite{o1,23}. However, simply locating each vehicle on the image is not enough for data preparation for semantic communication and multimodal-LLM. In the next phase, these parts and their surroundings are cropped from the image, and textual captions are generated for each of the cropped parts for multimodal-LLM tasks.  

\subsection{Vehicle Centric RoI Part Cropping and Text Captions}

YOLOv11 model provides us with the exact location of each vehicle in the camera images and the most essential parts that Multimodal-LLMs should focus on during query generation. However, two issues must be resolved before forwarding data to the multimodal-LLM. Visual-LLMs such as LLaVA are capable of handling 224x224 square size low-dimensional images \cite{24} while our camera captures 2048x2048 high-dimensional captions, which might cause the model to degrade in performance and increase the training and inference time during the query generations, which is improper for real-time applications. Moreover, although object detection model emphasizes the important points, redundant areas in the images still occupy the majority of the area, and these areas might impact the query of the multimodal-LLM since in real-time applications, it is essential to get a short and rapid query that highlights the important points.

Reducing camera captions to the proper dimensions might lead to a loss of essential features and make AI models work much more challenging. Therefore, another method, cropping the ROI points from the overall images, is applied to the camera captures. In this method, the rectangular bounding box collected by the YOLOv11 model is transformed into a square bounding box centered on the vehicle, and this square is extended with an $S$ scaling factor to see the vehicle's surroundings.

There are two different reasons for selecting the square size conventions. First, the size of each vehicle's rectangular bounding boxes varies depending on the vehicle's position on the image, and the dataset of the AI models, especially images, requires fixed-dimensional data. However, since vehicles' rectangles can have different widths and heights, trying to reshape them into the same dimension results in changes in the aspect ratio of the images and a loss of visual consistency. On the other hand, converting them into a square shape preserves the object’s central alignment and ensures consistent input shape. 

The second reason is making it easier to fit different squares into the same dimension, which is proper for the LLaVA model. One of the most proper image shapes for the LLaVA model is a 224x224 square size. Although we converted each vehicle's bounding boxes into square boxes, their dimensions can vary and be different from the required size. Fortunately, the square size enables the dimension changes to upper or lower size with minimal aspect ratio loss. It is also mentioned that we are collecting not only the vehicle's square size bounding box but also its $S$ scaling surroundings so that it will remove the redundant areas and provide the most essential parts for this vehicle activity. The algorithmic flow can be seen in the pseudocode in Alg. \ref{alg:square-bbox}

\begin{algorithm}
\caption{Adjust to Square Bounding Box} \label{alg:square-bbox}
\fontsize{7.5}{7.7}\selectfont
\KwIn{Bounding box $(x_1, y_1, x_2, y_2)$, image dimensions $(W, H)$, scale factor $s$}
\KwOut{Square bounding box $(x'_1, y'_1, x'_2, y'_2)$}

$bb_w \gets x_2 - x_1$, \quad $bb_h \gets y_2 - y_1$\;
$new\_w \gets bb_w \cdot s$, \quad $new\_h \gets bb_h \cdot s$\;
$L \gets \max(new\_w, new\_h)$\;
$cx \gets x_1 + bb_w / 2$, \quad $cy \gets y_1 + bb_h / 2$\;
$x'_1 \gets cx - L / 2$, \quad $x'_2 \gets cx + L / 2$\;
$y'_1 \gets cy - L / 2$, \quad $y'_2 \gets cy + L / 2$\;

\If{$x'_1 < 0$}{
    $x'_2 \gets x'_2 - x'_1$\;
    $x'_1 \gets 0$\;
}
\If{$y'_1 < 0$}{
    $y'_2 \gets y'_2 - y'_1$\;
    $y'_1 \gets 0$\;
}
\If{$x'_2 > W$}{
    $shift \gets x'_2 - W$\;
    $x'_1 \gets x'_1 - shift$\;
    $x'_2 \gets W$\;
}
\If{$y'_2 > H$}{
    $shift \gets y'_2 - H$\;
    $y'_1 \gets y'_1 - shift$\;
    $y'_2 \gets H$\;
}
\Return $\lfloor x'_1 \rfloor, \lfloor y'_1 \rfloor, \lfloor x'_2 \rfloor, \lfloor y'_2 \rfloor$
\end{algorithm}

\begin{figure*}
    \centering
    \includegraphics[width=0.8\linewidth]{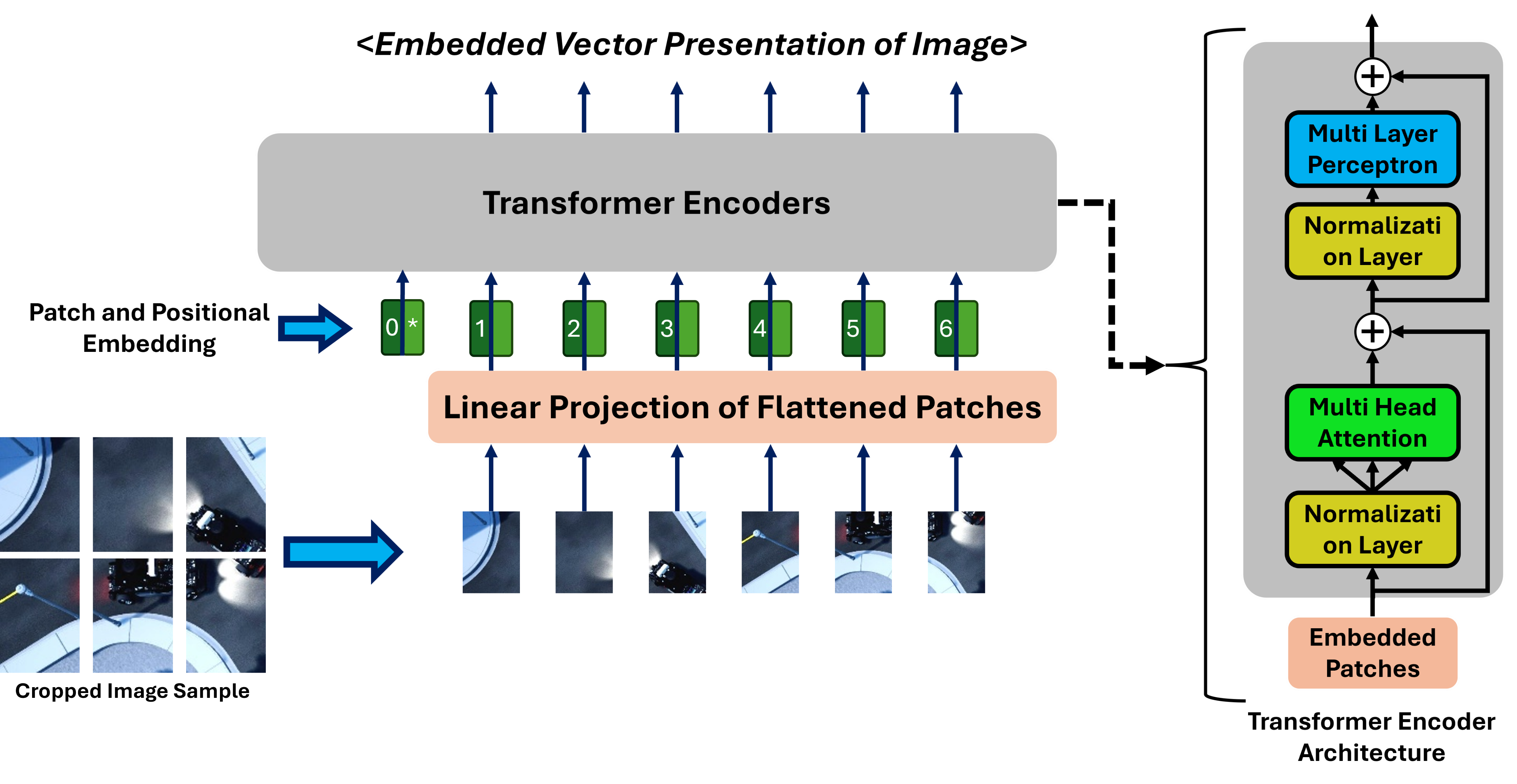}
    \caption{Vision Transformer Model and Transformer Encoder Architecture \cite{22}}
    \label{fig:vit}
\end{figure*}

According to Alg. \ref{alg:square-bbox}, the method takes rectangle bounding boxes $<x1, y1, x2, y2>$ coordinates, which are collected from the YOLOv11 model, as the first input. The second input is the overall image dimension, and the last input is the scale factor $S$.  The main reason for taking the overall image size is to check whether the rectangle bounding box is close to the edges or not. If it is close to the edge, the center of the square size changes due to taking not only the vehicle but also its surroundings with the $S$ factor. Otherwise, the center of the square remains as the vehicle's center location, and these extended bounding box coordinates are returned as an output.

In the Alg. \ref{alg:square-bbox}, initially, the width and height of the original bounding box are calculated using the edge coordinates $<x1, y1, x2, y2>$ (step 1). Then these width and height are scaled by the multiplication with the scale factor $S$ to get the vehicle's surroundings (step 2). Since a square size bounding box is required, the edge length, $L$ is set to the maximum of the scaled width and height (step 3). After that, the central location of x and y of the vehicle is calculated by using the original coordinates and original width and height (step 4). This central location and $L$ are used to get new $<x'_1, x'_2, y'_1, y'_2>$ coordinates of a square-sized scaled bounding box (step 5-6). The following steps (7-24) check whether the new coordinates exceed the original image edges or not by using the overall image width ($W$) and height ($H$), and if they do, the new coordinates are shifted into proper positions. After cropping each vehicle's part from the camera image, these parts are textually labeled manually for Multimodal-LLM training.

%In terms of computational complexity of methods in Alg. \ref{alg:square-bbox}, subtractions, multiplications, divisions, comparisons, \texttt{max}, and \texttt{floor} be $O(1)$ primitives. Algorithm~\ref{alg:square-bbox} performs a constant number of arithmetic operations to (i) compute the original box size $(bb_w, bb_h)$, (ii) scale to $(new\_w, new\_h)$, (iii) set the square side $L=\max(new\_w,new\_h)$, (iv) compute the center $(c_x,c_y)$, and (v) derive the square edges $(x'_1,y'_1,x'_2,y'_2)$. The subsequent boundary handling involves at most four conditional checks (left, top, right, bottom), each with constant work (a comparison and a few assignments; the right/bottom checks also include a constant-time shift). There are no loops or data-dependent iterations that grow with the image or box size. Therefore, the time complexity per bounding box is $\Theta(1)$ and the extra space complexity is $\Theta(1)$. For a frame with $N$ detections, the total complexity is $\Theta(N)$ time and $\Theta(1)$ extra space, and the per-box operations are independent and fully parallelizable.

Computational complexity of primitive operations in Alg. \ref{alg:square-bbox} such as additions, subtractions, multiplications, divisions, comparisons, \texttt{max}, and \texttt{floor} is $O(1)$. The overall step in Alg.~\ref{alg:square-bbox}, which computes $(bb_w,bb_h)$, applies scaling, determines $L=\max(\cdot)$, finds the center $(c_x,c_y)$, and clamps the square edges. Therefore, its per-box complexity is $O(1)$. After adjustment, each bounding box is extracted and resized to a fixed shape of 224×224 pixels from the original image. Since some bounding boxes might overlap with each other, the crop method is copying from the original camera captions, which requires $O(L_i^2)$ complexity per box, where  $L_i$ denotes the side length of the $i$-th square crop. In the next stage of the cropping of the bounding box, it is converted into a 224x224 fixed dimension. Computational complexity of image resizing is $O(X \cdot Y)$ where $X$ and $Y$ are the cropped edge sizes, regardless of the original image size. On the other hand, the extra space requirement is $O(X \cdot Y)$ to process patches sequentially,  Aggregating over $B_X$ boxes in a frame, the total complexity becomes:
\begin{equation}
T(B_X) = O\left(\sum_{i=1}^{B_X} L_i^2\right) + O\left(B_X \cdot X \cdot Y\right).
\label{eq:complexity-bbox}
\end{equation}

\subsection{Vision Transformers for Image Vector Embedding Generation}

In the previous section, the ROIs in the images, mostly vehicles and their surrounding areas, are extracted and resized into a \(224\times224\) square shape. Each cropped part is then textually labeled, completing the multimodal-LLM dataset generation process. However, transmitting all cropped images to the cloud can be inefficient, especially when multiple vehicles appear simultaneously in the camera view, as the communication cost scales with the number of detections. To reduce transmission overhead, we instead convert each cropped image into a compact embedded vector representation, which can be reconstructed or directly processed at the receiving side.

For this purpose, we employ a Vision Transformer (ViT)~\cite{22}, which outputs a fixed-size embedding of dimension, \(1\times768\), for each cropped image. ViT uses a transformer encoder as its backbone, leveraging the self-attention mechanism to capture both local details and global contextual relationships between different parts of an image~\cite{4}. While CNN-based feature extractors excel at learning local patterns, they rely on progressively increasing receptive fields to model global dependencies, which may limit their ability to capture long-range relationships in complex scenes such as traffic environments with multiple interacting objects. In contrast, ViT’s self-attention mechanism computes relationships between all patch pairs in a single layer, allowing the model to represent spatially distant yet semantically related objects more effectively.

The ViT input pipeline converts each image into a sequence of non-overlapping patches, treating each patch analogously to a token in NLP tasks. These patches are flattened, linearly projected into vectors, and augmented with positional encodings to retain spatial layout information. This sequence of patch tokens is then processed by multiple transformer encoder layers, where multi-head self-attention adaptively weights the contribution of each patch to the final embedding.

\subsubsection{{ViT Embeddings Generation}}

Given an input image $I \in \mathbb{R}^{H \times W \times C}$ with height $H$, width $W$, and $C$ color channels, the ViT first splits the image into $P$ non-overlapping patches of size $p \times p$, where 
$P = (H \times W)/{p^2}$.
Each patch is flattened into a vector $x_i \in \mathbb{R}^{p^2 \cdot C}$ and projected into a $D$-dimensional space using a learnable linear projection $\mathbf{E} \in \mathbb{R}^{(p^2 \cdot C) \times D}$, resulting in 
\begin{equation}
\mathbf{z}_i^0 = x_i \mathbf{E}, \qquad \forall i \in \{1,\dots,P\}.
\end{equation}

A learnable class token $\mathbf{z}_{\text{cls}}^0 \in \mathbb{R}^D$ is then placed at the beginning of the sequence of patch embeddings, and learnable positional embeddings $\mathbf{E}_{\text{pos}} \in \mathbb{R}^{(P+1) \times D}$ are added to retain spatial order information. The resulting input sequence becomes
\begin{equation}
\mathbf{Z}^0 = [\mathbf{z}_{\text{cls}}^0; \mathbf{z}_1^0; \dots; \mathbf{z}_P^0] + \mathbf{E}_{\text{pos}}.
\end{equation}

This token sequence is passed through $L_s$ stacked Transformer encoder blocks, each consisting of a Multi-Head Self-Attention (MSA) module and a Feed-Forward Network (FFN). 
%At each layer $\ell$, the computation is given by}
% \[
% \mathbf{Z}^{\ell'} = \text{MSA}\big(\text{LN}(\mathbf{Z}^{\ell-1})\big) + \mathbf{Z}^{\ell-1},
% \qquad
% \mathbf{Z}^{\ell} = \text{FFN}\big(\text{LN}(\mathbf{Z}^{\ell'})\big) + \mathbf{Z}^{\ell'}.
% \]

After the final encoder block, the output corresponding to the class token is extracted as the global image representation:
\begin{equation} 
\mathbf{h} = \mathbf{z}_{\text{cls}}^{L_s} \in \mathbb{R}^D.
\end{equation}
Figure~\ref{fig:vit} illustrates the adopted ViT-based embedding extraction process.

%In our edge–cloud architecture, this approach offers two main benefits: (i) it produces semantically rich and fixed-size embeddings that require significantly less bandwidth than transmitting raw cropped images, and (ii) it retains the full contextual understanding of the vehicle’s visual surroundings, which is critical for downstream tasks such as traffic behavior analysis and multimodal reasoning in the cloud. Figure~\ref{fig:vit} illustrates the adopted ViT-based embedding extraction process.

\subsection{System Model for Wireless Transmission of Embeddings}

This subsection presents a system model for wireless transmission for vector embeddings, with a focus on transmitting image embeddings over an Additive White Gaussian Noise (AWGN) channel. The system includes: (i) bitstream encoding of float32 type vector embeddings using both uniform quantization and IEEE 754  format, (ii) digital modulation using BPSK and 16-QAM, (iii) signal transmission over an AWGN channel, (iv) demodulation, and (v) reconstruction of the vector embeddings at the receiver. The system performance is evaluated using Mean Squared Error (MSE) between the transmitted and reconstructed embeddings.

\subsubsection{Bitstream Encoding}

Given a vector embedding:
\begin{equation}
\mathbf{v} = [v_1, v_2, \dots, v_d]^\top, \quad v_i \in \mathbb{R},
\end{equation}
we employ two encoding schemes: First is Uniform Quantization, where each $v_i$ is quantized to an $n_q$-bit integer value:
\begin{equation}
q_i = \left\lfloor \frac{v_i - v_{\min}}{v_{\max} - v_{\min}} \cdot (2^{n_q} - 1) \right\rfloor
\end{equation}
Here, $q_i \in {0, 1, ..., 2^{n_q} - 1}$ denotes the $i$-th quantized value, and $v_{\min}$ and $v_{\max}$ are the minimum and maximum bounds of the original embedding distribution, respectively.

The second encoding scheme is IEEE 754 Format, where each float32 type value, V, of embedding is converted into a 32-bit binary representation using the IEEE 754 standard, consisting of 1 sign bit, S, 8 exponent bits, E, and 23 fraction bits, M, also referred to as the mantissa. Using these components, the value of embedding is represented as:
\begin{equation}
 V= (-1)^S \times 2^{(E - 127)} \times (1 + M).
 \end{equation}

The resulting bitstream from all $d$ elements is:
\begin{equation}
\mathbf{b} = [b_1, b_2, \dots, b_{Nd}] \in \{0, 1\}^{Nd}
\end{equation}
where $N = 32$ for IEEE 754 or $N = n_q$ for uniform quantization encoding.

\begin{table}[t]
\caption{Notation used in Section III-E }
\label{tab:notation-III-E}
\centering
\renewcommand{\arraystretch}{1.15}
\begin{tabular}{ll}
\toprule
\textbf{Symbol} & \textbf{Meaning} \\
\midrule
$\mathbf{v}=[v_1,\ldots,v_d]^\top$ & Embedding vector of dimension $d$ \\
$d$ & Dimensionality of the embedding vector \\
$v_i$ & $i$-th element of the embedding vector \\
$v_{\min},\, v_{\max}$ & Min/Max bounds used for uniform quantization \\
$q_i$ & Quantized integer of $v_i$  \\
$n_{\text{q}}$ & Quantization bit-depth (e.g., $8,16,32$) \\
$N$ & Bits per element  \\
$S$ & IEEE 754 sign bit \\
$E$ & IEEE 754 exponent (8-bit) \\
$M$ & IEEE 754 fraction/mantissa (23-bit) \\
$V$ & IEEE 754 value\\
$\mathbf{b}\in\{0,1\}^{Nd}$ & Bitstream formed from all $d$ elements \\
$b,\, b_i$ & A single bit / the $i$-th bit in the bitstream \\
$s$ & Modulated symbol  \\
$I,\,Q$ & In-phase and quadrature components (Gray coded) \\
$j$ & Imaginary unit, $j^2=-1$ \\
$\mathbf{s}$, $\mathbf{y}$ & Sequence of transmitted and Received symbols \\
$\mathbf{n}\sim\mathcal{N}(0,\sigma^2)$ & AWGN noise samples \\
$E_s$ & Average symbol energy \\
$\mathrm{SNR}$ & Signal-to-noise ratio in dB \\
$\sigma^2$ & Noise variance for a given SNR \\
$\hat{b}$ & Hard-decision demodulated bit (BPSK) \\
$\hat{I},\,\hat{Q}$ & Nearest-neighbor decisions for I/Q (QAM16) \\
$\Re(\cdot),\,\Im(\cdot)$ & Real and imaginary part operators \\
$\hat{v}_i$ & Reconstructed $i$-th embedding element \\
$\hat{\mathbf{v}}$ & Reconstructed embedding vector \\
\bottomrule
\end{tabular}
\end{table}

\subsubsection{Modulation Scheme and Communication Channel}
We have considered two modulation schemes: Binary Phase shift Keying (BPSK) and 16-Quadrature Amplitude Modulation (16-QAM). In BPSK modulation, each bit $b \in \{0,1\}$ is mapped to a real-valued symbol $s \in \{-1, +1\}$:
\begin{equation}
s = 2b - 1.
\end{equation}
In the case of 16-QAM modulation,
groups of 4 bits are mapped to complex symbols using Gray coding. Let $b_i$ be the $i$-th bit in group:
\begin{equation}
[b_1, b_2, b_3, b_4] \rightarrow s = I + jQ, \quad I, Q \in \{-3, -1, +1, +3\}
\end{equation}
The modulated signal $\mathbf{s}$ is transmitted over an AWGN channel:
\begin{equation}
\mathbf{y} = \mathbf{s} + \mathbf{n}, \quad \mathbf{n} \sim \mathcal{N}(0, \sigma^2)
\end{equation}
where the Gaussian noise variance $\sigma^2$ depends on the desired SNR:
\begin{equation}
\sigma^2 = \frac{E_s}{10^{\text{SNR}/10}}
\end{equation}

At the receiver side, for BPSK demodulation,
the demodulated bit is:
\begin{equation}
\hat{b} = \begin{cases} 1 & \text{if } y \geq 0 \\ 0 & \text{otherwise} \end{cases}
\end{equation}
For 16-QAM Demodulation, each received complex symbol is mapped to the nearest constellation point $(i+jq)$:
\begin{align}
\hat{I} &= \arg\min_{i \in \{-3,-1,1,3\}} |\Re(y) - i| \\
\hat{Q} &= \arg\min_{q \in \{-3,-1,1,3\}} |\Im(y) - q|
\end{align}
These are then mapped back to the original 4-bit groups using inverse Gray coding.
\subsubsection{Bitstream to Embedding Reconstruction}
The received bitstream is first parsed and decoded into its constituent embedding representations. 
If the bitstream represents embeddings quantized using uniform quantization over $n_q$ bits, then each integer $q_i$ in the bitstream is mapped back to its approximate floating-point value $\hat{v}_i$ using the following linear inverse transformation:
\begin{equation}
\hat{v}_i = \frac{q_i}{2^{n_q} - 1} (v_{\max} - v_{\min}) + v_{\min}
\end{equation}
This reconstruction recovers an approximation embedding vector, $\hat{\mathbf{v}}$, of the original embedding vector, ${\mathbf{v}}$, by interpolating within the known value range.

Alternatively, if each element in the bitstream is a 32-bit IEEE 754 representation of a floating-point value, the reconstruction is done using a binary-to-float conversion that follows the IEEE 754 standard. These are decoded to yield the float ${\hat{v}}_i = (-1)^S \times 2^{(E - 127)} \times (1 + M)$. This decoding is precise if the original values are encoded using the IEEE 754 format, ensuring full precision recovery.

While cosine similarity is often used for vector comparison, it is scale-invariant and may mask significant magnitude deviations caused by quantization or channel noise. In our framework, both the direction and magnitude of the embeddings influence the downstream image decoder and LLM performance. Therefore, to assess the quality of reconstruction, we compute the Mean Squared Error (MSE) between the original embedding vector $v$ and its reconstructed version $\hat{v}$:
\begin{equation}
\text{MSE} = \frac{1}{d} \sum_{i=1}^{d} (v_i - \hat{v}_i)^2
\end{equation}
where $v_i$ and $\hat{v}_i$ are the $i$-th original and reconstructed embedding values respectively, and $d$ is the dimensionality of the embedding vector.
A lower MSE indicates higher fidelity and more accurate reconstruction, which is critical for downstream tasks such as semantic understanding, classification, or retrieval. While cosine similarity is often used for vector comparison, it is scale-invariant and may mask significant magnitude deviations caused by quantization or channel noise. In our framework, both the direction and magnitude of the embeddings influence the  decoder and LLM performance. 

\subsection{Image Decoding on The Cloud Server} 

Image vector embeddings generated by ViT are transmitted to the cloud server using much less memory and bandwidth over a wireless channel. However, multimodal-LLMs require image data to generate a proper query for the traffic environment. Therefore, it is required to convert these vector embeddings to the original image form, which is 224x224 RGB images. For this purpose, on the cloud server, after receiving vector embeddings, these are sent to the image decoder model, which consists of a transpose CNN layer that is used to upscale feature maps in neural networks. 

\begin{figure*}
    \centering
    \includegraphics[width=0.85\linewidth]{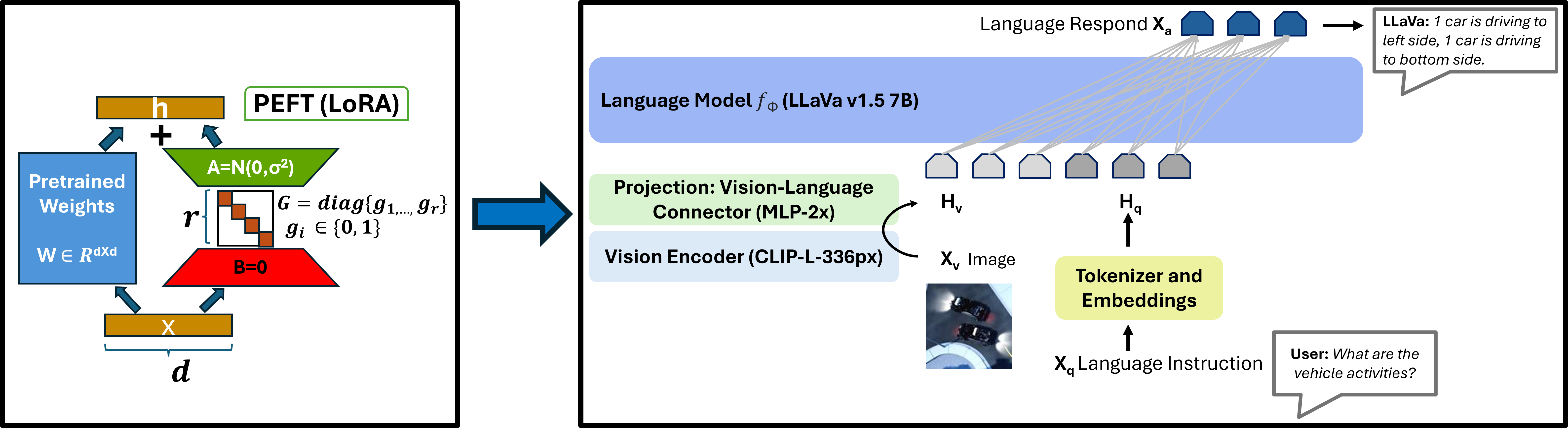} \vspace{0.2cm}
    \caption{Large Language and Vision Assistant (LLaVA) Model (on the Right) and LoRA Fine-Tuning (on the Left) \cite{24}}
    \label{fig:llava}
\end{figure*}

%At the beginning of the image decoding. Two-dimensional embedded vectors are unflattened into three dimensions and transposed CNN layers upscale these features into 224x224x3-dimensional images with a kernel size of 4 and a stride of 2. Learned Perceptual Image Patch Similarity (LPIPS) metric \cite{lpips}, which is a perceptual loss function designed to compare images based on how humans perceive visual differences, rather than relying solely on pixel-level or structural similarity, is used as a criterion function to compare the differences between original images and the reconstructed one from the decoder. In this research, LPIPS is calculated by passing image patches through a pre-trained CNN model, such as AlexNet, and computing the differences between the deep feature representations of the two images. 

At the beginning of the image decoding. Two-dimensional embedded vectors are unflattened into three dimensions and transposed CNN layers upscale these features into 224x224x3-dimensional images with a kernel size of 4 and a stride of 2. To quantify the perceptual quality of the reconstructed images, we employ the Learned Perceptual Image Patch Similarity (LPIPS) metric~\cite{lpips}, which measures the similarity between two images in a deep feature space rather than directly in the pixel space. Conventional pixel-wise metrics such as MSE or PSNR often fail to correlate well with human visual perception, as they treat all pixel errors equally without considering structural or semantic differences. LPIPS addresses this limitation by computing distances in the feature space of a pretrained CNN, such as AlexNet, which has been trained on large-scale image datasets, like ImageNet, and captures high-level perceptual representations.

Given a reconstructed image $\hat{\mathbf{x}}$ and a ground-truth image $\mathbf{x}$, both are passed through a fixed pretrained network $\phi$ to obtain a set of normalized feature maps $\{\hat{\mathbf{f}}_l, \mathbf{f}_l\}$ at multiple layers $l$. The LPIPS score is then computed as:
\begin{equation}
\text{LPIPS}(\mathbf{x}, \hat{\mathbf{x}}) = \sum_{l} w_l \cdot \frac{1}{H_l W_l} \sum_{h=1}^{H_l} \sum_{w=1}^{W_l} \left\| \mathbf{f}_l^{\,hw} - \hat{\mathbf{f}}_l^{\,hw} \right\|_2^2,
\label{eq:lpips}
\end{equation}
where $H_l$ and $W_l$ are the spatial dimensions of the $l$-th feature map, and $w_l$ is a learned weight reflecting the relative importance of each layer. A lower LPIPS score indicates higher perceptual similarity. In this framework, LPIPS is incorporated into the loss function to encourage reconstructions that are perceptually closer to the ground truth, aligning more closely with human visual judgments than pixel-wise error metrics. The choice of the pretrained backbone can influence sensitivity to different perceptual attributes, and we adopt AlexNet in our implementation.

\subsection{Large Language and Vision Assistant Model For Image Queries} \label{llm}

In the final step, after image reconstruction from the decoder, the images are forwarded to the multimodal-LLM to generate proper queries for urban traffic monitoring and management. Choosing the appropriate multimodal-LLM, which provides the best and efficient results, and is easy to implement and fine-tune the new custom dataset, is essential for real-time traffic queries.  We select LLaMA 3.2-11B Vision-Instruct \cite{meta2024llama} and LLaVA 1.5 7B \cite{24} for comparison since both are recent open-weight multimodal large language models capable of jointly processing visual inputs such as camera-based traffic images, and textual descriptions, making them directly applicable to edge–cloud traffic management scenarios. LLaMA 3.2-11B Vision-Instruct represents a state-of-the-art instruction-tuned vision–language model with a large 11B parameters for high-quality reasoning, while LLaVA 1.5 7B is a smaller, widely used multimodal model with lower computational requirements, making it suitable for resource-constrained environments. LLaVA 1.5 with 7B and LLaMA 3.2 with 11B parameters are compared in table \ref{tab:llama_vs_llava}.

%LLaMA is one of the well-known LLMs, and in recent studies, with LLaMA-3.2, model abilities are extended to process visual inputs alongside textual prompts \cite{meta2024llama}. On the other hand, LLaVA is another vision-language instruction-tuned model that integrates a pretrained LLM with a vision encoder to enable multimodal understanding \cite{24}. LLaVA 1.5 with 7B and LLaMA 3.2 with 11B parameters, which are their lightest versions, are compared in table \ref{tab:llama_vs_llava}.

\begin{table}[] 
\renewcommand{\arraystretch}{1.25}  
\centering
\caption{Comparison between LLaMA 3.2-11B Vision-Instruct and LLaVA 1.5 7B\\}
\label{tab:llama_vs_llava}
\begin{tabular}{lll}
%\textbf{}                                                                                 & \multicolumn{1}{c}{\textbf{\begin{tabular}[c]{@{}c@{}}LLaMA   3.2-11B \\ Vision-Instruct\end{tabular}}} & \multicolumn{1}{c}{\textbf{LLaVA   1.5 7B}}                                                  \\
\textbf{Feature} &   \textbf{LLaMA 3.2-11B}  & \textbf{LLaVA 1.5 7B}                                                                                             \\ \hline
\textbf{Language   Backbone}                                                              & LLaMA 3.1                                                                                       & Vicuna 1.5                                                                       \\ \hline
\textbf{Vision   Backbone}                                                                & \begin{tabular}[c]{@{}l@{}}Native vision  \\ model from Meta\end{tabular}                               & \begin{tabular}[c]{@{}l@{}}CLIP ViT-L/14   or ViT-H/14 \\ (from OpenAI/LAION)\end{tabular}   \\ \hline
\textbf{Projection   Layer}                                                               & \begin{tabular}[c]{@{}l@{}}Internal \\ visual adapter\end{tabular}                                      & \begin{tabular}[c]{@{}l@{}}Custom MLP\\ projector (mlp2x\_gelu)\end{tabular}                 \\ \hline
\textbf{Multimodal   Fusion}                                                              & \begin{tabular}[c]{@{}l@{}}Unified  \\ architecture\end{tabular}                                        & \begin{tabular}[c]{@{}l@{}}External vision-language \\ alignment via projection\end{tabular} \\ \hline
\textbf{VRAM   Efficiency}                                                                & 24 GB                                                                                                   & 14.2 GB                                                                                      \\ \hline
\textbf{Model   Size (Disk)}                                                              & 7.9GB                                                                                                   & 4.7GB                                                                                        \\ \hline
\textbf{\begin{tabular}[c]{@{}l@{}}Inference   Time \\ (1 image + 1 prompt)\end{tabular}} & $\sim$3.5–4.5 sec                                                                                       & $\sim$0.8–1.5s                                                                               \\ \hline
\end{tabular}
\end{table}

The main reason for selecting these two models is to cover two distinct but representative ends of the spectrum: (i) a high-capacity model prioritizing accuracy and reasoning ability, and (ii) a lighter model with reduced inference cost that generates brief responses in a short time. Meta's LLaMA 3.2-11B Vision-Instruct is a multimodal LLM that integrates visual understanding into the LLaMA 3.1 architecture. The model includes 11 billion parameters and is designed for tasks such as image captioning, visual question answering (VQA), and image-text reasoning. The model utilizes a vision adapter with cross-attention layers to integrate the image encoder's outputs into the language model, allowing it to process both text and images. With a context window of up to 128k tokens, it supports high-resolution multimodal inputs. Training involved 6 billion image-text pairs, and while it supports multiple languages for text-only tasks.

{On the other hand, LLaVA 1.5 7B is an open-source multimodal-LLM that uses the Vicuna-1.5 as a base model and incorporates the CLIP ViT-L/14-336px vision encoder. It utilizes a two-layer multi-layer perceptron (MLP) vision-language connector, known as mlp2x\_gelu, to align visual and textual representations. Trained on GPT-4-generated multimodal instruction-following data, LLaVA 1.5 demonstrates strong performance in tasks like VQA and image captioning. 

However, when comparing LLaMA and LLaVA models with each other, the LLaMA model requires 7.9 GB of disk space, while this requirement is 4.7 GB in LLaVA 1.5. Moreover, in order to run LLaMA with proper efficiency, 24 GB VRAM is required, and it is 14.2 GB in the LLaVA 1.5 model. Furthermore, LLaMA inference times for images, which are around 3.5-4.5 seconds, are higher than LLaVA 1.7, which is around 0.8-1.5 seconds. Although LLaMA is trained with more tokens and can provide more detail for images, the LLaVA model, due to being the lightest LLaMA-based model that processes images with minimum inference time and requires less memory, is suitable for real-time traffic query generation.

Fine-tuning LLaVA on custom datasets (See in Section \ref{dataset}) is notably efficient due to its modular architecture and compatibility with parameter-efficient techniques like Low-Rank Adaptation (LoRA) and DeepSpeed. Training with LoRA and Architecture of LLaVA 1.5 is illustrated in \figurename \ref{fig:llava}.

In \figurename \ref{fig:llava}, the left part shows the LoRA fine-tuning part. LoRA is a method that is used to adapt LLMs to specific use cases by adding lightweight pieces to the original model rather than changing the entire model. In normal circumstances, training massive machine learning models, which include billions of parameters, for specific contexts requires a great deal of retraining that takes a long time and requires numerous resources such as data or memory. On the other hand, LoRA provides a quick way to adapt the model without retraining it. This method freezes the initial, pretrained model parameters and weights rather than retraining the entire model. Then, it applies a low-rank matrix, which is a lightweight modification to the existing model, to fresh inputs to produce context-specific outputs. The low-rank matrix adjusts for weights of the original model so that outputs match the desired use case.

On the other hand, the right part of \figurename \ref{fig:llava} depicts the architecture of LLaVA v1.5 7B multimodal-LLM that integrates both visual and textual inputs. LLaVA 1.5 7B model derived from Vicuna-1.5, which is a fine-tuned version of LLaMA 2 7B and performs joint reasoning across image and text by aligning the visual tokens within the token stream of the language model. It uses OpenAI’s CLIP Vision Transformer Large (ViT-L/14) as a vision encoder to convert the raw input images into a sequence of visual embeddings. Moreover, a two-layer MLP with GELU activations, which is referred to as mlp2x\_gelu, is integrated to bridge the gap between image embeddings, from CLIP, and language models' input space by projecting the high-dimensional visual tokens into a space compatible with the text model’s embedding dimensions. The flow of this model illustrates that initially, images, $X_{v}$, are processed by the CLIP encoder and transformed into patch embeddings. These embeddings are then aligned with the text embedding space, $H_{q}$, generated by tokenization and embedding the textual data, $X_{q}$,  by the mlp2x\_gelu projector, $H_{v}$. All tokens, $H_{q}$ and $H_{v}$, are forwarded to the LLaVA model, and the language response is generated.

\section{Training and Numerical Results} \label{sec:results}

\begin{figure}
    \centering
    \includegraphics[width=0.8\linewidth]{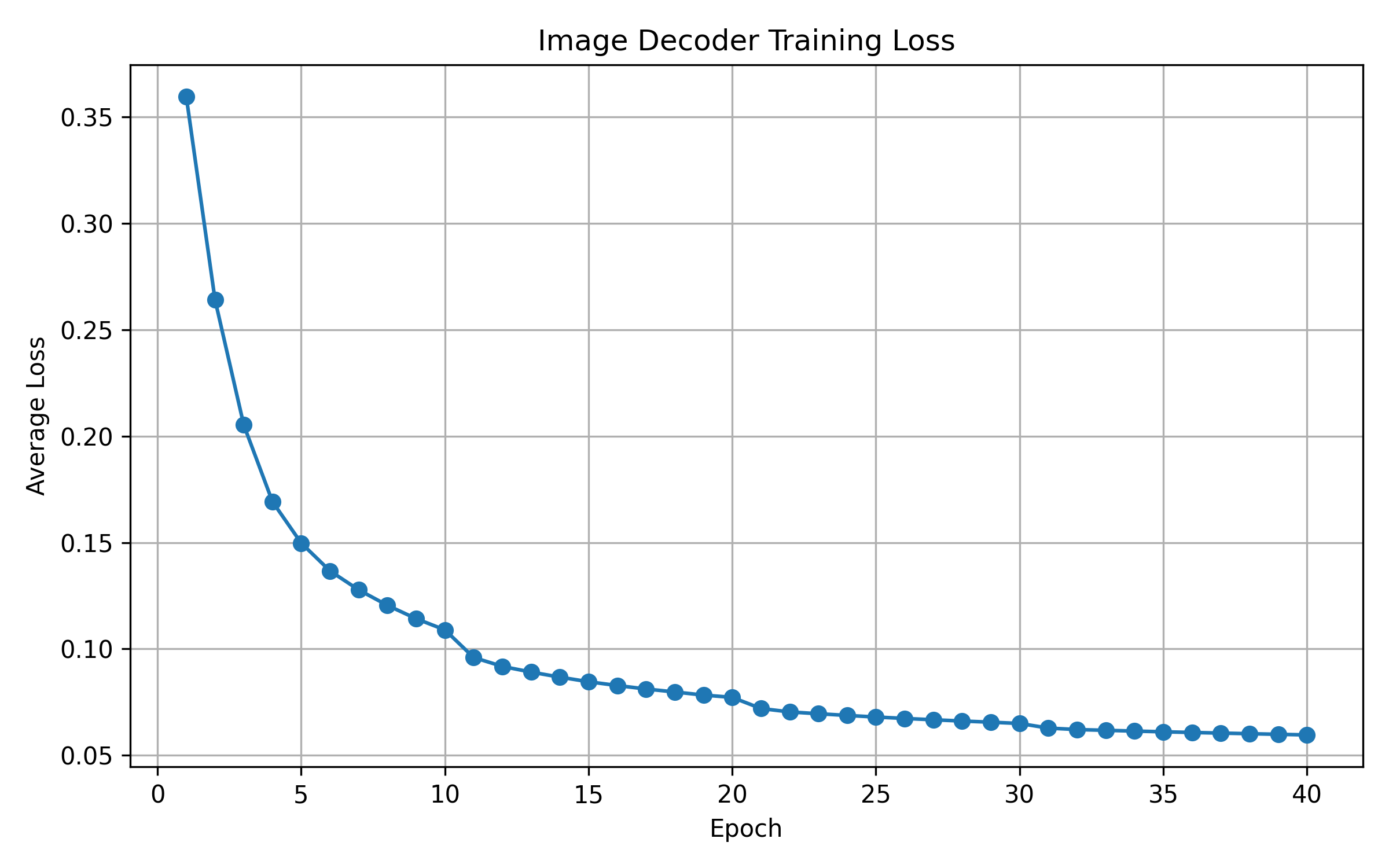}\vspace{0.1cm}
    \caption{Convergence Plot of Image Decoder in 40 Epochs}
    \label{fig:convergence_decoder}
\end{figure}

\begin{figure*}
    \centering
    \includegraphics[width=0.8\linewidth]{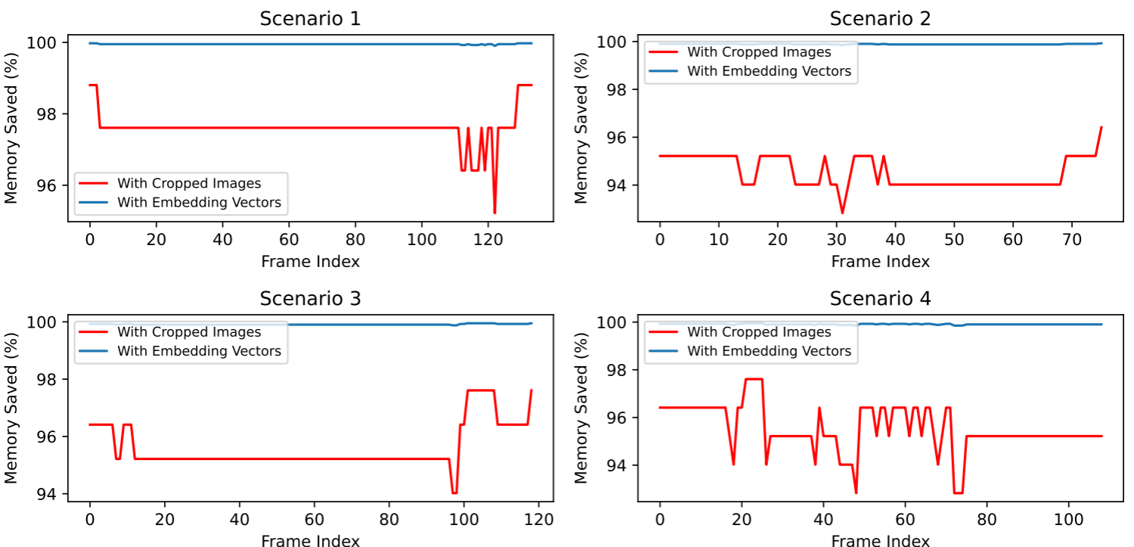}\vspace{0.1cm}
    \caption{Memory Saving with Cropped Images and with Embedded Vectors}
    \label{fig:results}
\end{figure*}

\subsection{Training Setup and Parameters}
In the framework setup, a pre-trained YOLOv11 model is deployed from our previous work \cite{onsu2025leveraging} for vehicle detection during the simulation runtime. Moreover, already pre-trained variants of ViT, \textit{vit-base-patch16-224-in21k}, are integrated into the proposed framework for image embedding vector representation. For LLaVA 1.5 7B and image decoder model training, Qlab simulation runs 20 times with various scenarios, and the cropping essential blocks phase starts. During the cropping phase, the scaling factor, $S$, is a crucial parameter since it allows the multimodal-LLM to see not just the vehicle but also its surroundings. Deciding on the scaling factor is critical, as it restricts the multimodal-LLMs' sight to only the vehicle itself if it is smaller, such as $2.0$. On the other hand, if it is larger, such as $3.0$, it makes the process challenging due to the redundant background information and the increased image size. Therefore, the scaling factor is chosen as $2.5$ to balance the trade-off between capturing sufficient contextual surroundings and avoiding the excessive irrelevant background. After bounding box coordinates of each vehicle are collected from the YOLO detector. After its center point is computed, we determine the maximum of its width or height as the base edge length. Multiplying this length by $2.5$ ensures that the cropped region covers the object together with nearby road markings, adjacent vehicles, and relevant environmental cues. %After some experiments, the best scaling factor is determined to be 2.5 times the longest edge of the bounding box of each vehicle. 
20 different longer scenarios are established in Qlab which provides diverse high quality dataset such as roundabouts, intersections, different speed vehicles, collisions and so on, and every 5 ms new image data is collected from the scenario runtime. Among all these experiments, by utilizing Alg. \ref{alg:square-bbox}, 2400 cropped images are collected from overall camera data. Each cropped images includes vehicles in its center location, and these are passed through a ViT to extract embedded vectors, which are stored in a database along with the original images and captions. In the training  phase, a custom decoder reconstructs images by expanding these embeddings through a fully connected input layer into high-dimensional feature maps. This feature map is unflattened into three-dimensional data and then progressively up-sampled through a series of five ConvTranspose2d layers, each of which is followed by a ReLU activation to gradually increase spatial resolution while reducing the number of feature channels. The final transposed convolution layer outputs a 3-channel image, and a Sigmoid activation scales pixel values to the range [0, 1], suitable for image reconstruction tasks. LPIPS is used as a loss function, and the model is trained with 40 epochs and the loss value between the original image and the reconstructed image once training is finalized is 0.0549, which convergence plot can be seen in \figurename  \ref{fig:convergence_decoder}. After the image Decoder model training is concluded, reconstructed images, alongside their textual captions, are sent to the LLaVA 1.5 7B model training, and the LoRA fine-tuning approach is utilized to train a model on our specific traffic dataset with 20 epochs which is decided from our previous works related to Intelligent Multimodal-LLM based Traffic monitoring \cite{onsu2024leveraging}.  Training Setup and parameters can be seen in Table \ref{tab:setup}.

\begin{table}[]
\renewcommand{\arraystretch}{1.25}  
\centering
\caption{Training Setup and Parameters}
\label{tab:setup}
\begin{tabular}{ll}
                       & \textbf{Detail}            \\ \hline
YOLO                   & YOLOv11                    \\ \hline
Scaling Factor         & 2.5                        \\ \hline
ViT                    & vit-base-patch16-224-in21k \\ \hline
Image Decoder          & 5 ConvTranspose2d+ReLU     \\ \hline
Decoder Loss           & LPIPS                      \\ \hline
LLaVA                  & LLaVa 1.5 7B               \\ \hline
\#Scenarios            & 20                         \\ \hline
\#Samples              & 2400 Cropped Images        \\ \hline
\#Epochs Image Decoder & 40                         \\ \hline
\#Epochs LLaVA         & 20                         \\ \hline
\end{tabular}
\end{table}

\subsection{Numerical Results}
The embedding vectors generated at the edge are transmitted to the cloud over a wireless AWGN channel using two encoding schemes: IEEE 754 standard and Quantization. The evaluation includes various performance metrics such as communication efficiency in terms of bandwidth savings achieved by reducing transmission data size, mean squared error (MSE) of reconstructed embeddings across different encoding and modulation schemes, visual comparison and LPIPS scores of the reconstructed images at the receiver, and the accuracy of responses generated by the multimodal LLM. 

\figurename \ref{fig:results} represents the memory saving, or in other words, reduction in transmission data size in terms of percentages, with both cropped images and image embedding vector representations. The original image captured by the camera has a 2048x2048 RGB size, which is around 12.01 MB. On the other hand, each RoI area, cropped from the overall images is 224x224 RGB square image form that consumes around 0.14 MB, so the total memory consumption from 1 camera image, after cropping the highlighted parts, is $N$x0.14 MB, where $N$ is the total number of cars located by the YOLOv11 model. Furthermore, each image embedding representation is formed into a 1×768 float32 type vector data structure and occupies much less memory, around 29x$10^{-4}$ MB, and just like cropped images, the total memory requirement is $N$x29x$10^{-4}$
 MB for transmitting the embedding representation of one image frame to the cloud. In both cases, the overall amount of memory needed for transmission increases as the number of vehicles rises. However, cropped images and embedded vector representations occupy much less memory than the original images. For this reason, regardless of the number of vehicles in the camera views in  simulation, the total memory requirement doesn't exceed the original memory requirement in both cases.

%In the left part of \figurename \ref{fig:results}, memory saving using cropped images in four different scenarios with various durations is illustrated. The x-axis shows the frame index, which can be treated like a timestamp, and the y-axis is the memory savings for each frame in terms of percentage. The constant memory savings between indexes means that the number of vehicles during that duration is the same. If memory saving increases, it shows that vehicles are leaving the camera's view, and if memory consumption increases, it indicates that more vehicles enter the camera view. According to the results, minimum memory savings can be seen in the fourth scenario with 11.2 MB, which signifies the maximum amount of vehicles in that testing case, while maximum savings is around 11.9 MB in the first scenario. 

\figurename~\ref{fig:results} presents the percentage memory savings per frame across four different scenarios of different durations. The x-axis denotes the frame index, effectively representing the temporal progression of each scenario. The variation in memory savings among different scenarios for the cropped image method is due to differences in average vehicle count and relative ROI sizes per frame. For example, Scenario 1 contains fewer and generally smaller detected vehicles, leading to fewer 224×224 cropped images being transmitted, and thus slightly higher memory savings. In Scenario 2, higher traffic density and larger detected vehicles increase the total cropped image count per frame, resulting in a marginally lower savings percentage. Among the evaluated methods, the cropped image-based approach achieves memory savings ranging from approximately 93\% to a peak of 98.5\%. In contrast, converting cropped images into embedding vectors offers a consistent memory saving of around 99.9\% across all cases. While the number of vehicles in the camera's field of view influences memory usage in both approaches, the significantly smaller size of embedding vectors compared to cropped images leads to minimal variance in memory savings when using embeddings. This makes the vector-based method more stable and scalable in dynamic scenes.

Given that the framework operates in real-time, inference latency must also be considered. Building upon the previous pipeline in \cite{onsu2025leveraging}, which already employs YOLOv11 and LLaVA, the proposed system introduces three additional processing steps. For each frame, Region of Interest (RoI) extraction requires 0.023 seconds, embedding generation using a ViT takes 0.16 seconds, and image reconstruction through a custom decoder incurs a negligible latency of 1×$10^{-3}$ seconds. When using only the RoI cropping method, the total inference time scales as 0.023×$N$ seconds, where $N$ is the number of detected vehicles in one image frame. With the embedding-based method, the total inference time increases to 0.183×$N$ seconds, which includes both RoI extraction time and embedding generation time. Despite the added computational cost, the embedding-based approach yields superior memory efficiency, achieving up to 99.9\% data reduction during transmission. A comparative summary of the memory requirements for original images, cropped images, and embedding vectors is provided in Table~\ref{tab:mem}.
\begin{figure}
    \centering
    \begin{subfigure}[b]{0.35\textwidth}
        \includegraphics[width=\linewidth]{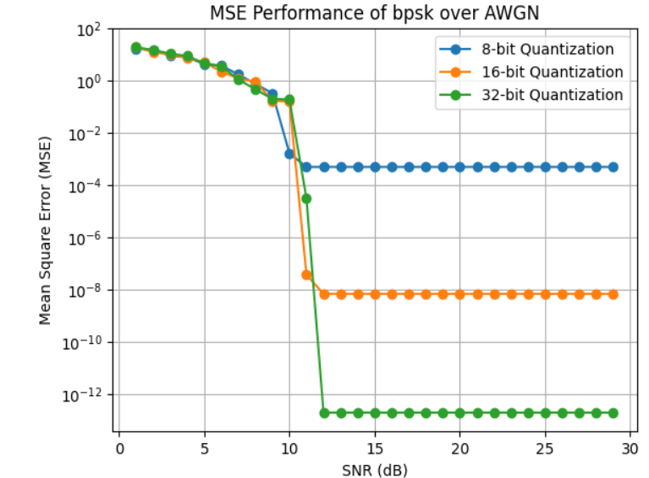}
       % \caption{Caption 1}
        \label{fig:sub1}
    \end{subfigure}
  \hfill
    \begin{subfigure}[b]{0.35\textwidth}
        \includegraphics[width=\linewidth]{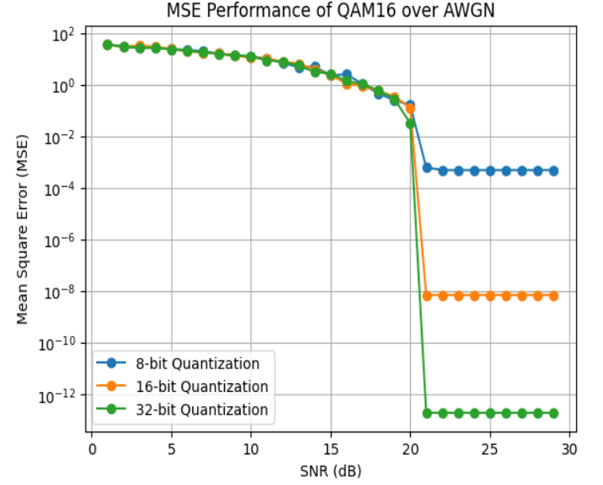}
       % \caption{Caption 2}
        \label{fig:sub2}
    \end{subfigure}
\caption{Comparison of MSE in reconstructed embeddings across different modulation schemes for Quantization encoding.}
    \label{fig:Quant}
\end{figure}

\begin{table}[]
\renewcommand{\arraystretch}{1.25}  
\centering
\caption{Memory Requirements For Each Data Type}
\label{tab:mem}
\begin{tabular}{llll}
                                                                                            & \textbf{Dimensions} & \textbf{Data Type} & \textbf{Memory Size} \\ \hline
Original Image                                                                              & 2048x2048x3         & Pixel              & 12.01 MB             \\ \hline
Cropped Image                                                                               & 224x224x3           & Pixel              & 0.143 MB             \\ \hline
\begin{tabular}[c]{@{}l@{}}Embedded Vector\\ Representation of\\ Cropped Image\end{tabular} & 1x768               & Float              & 29x$10^{-4}$ MB      \\ \hline
\end{tabular}
\end{table}

Figures~\ref{fig:Quant} and~\ref{fig:IEEE} illustrate the Mean Squared Error (MSE) performance of reconstructed embedding vectors transmitted over an AWGN channel under various modulation schemes (BPSK and QAM16) and encoding techniques (Quantization and IEEE 754). In Figures~\ref{fig:Quant}, the performance of Quantization encoding is evaluated using 8-bit, 16-bit, and 32-bit precision levels. The top plot shows results for BPSK modulation, while the bottom plot corresponds to QAM16 modulation. As expected, higher quantization precision leads to lower reconstruction error. Specifically, for BPSK, the MSE drops sharply as the SNR increases, particularly beyond 10 dB. 32-bit quantization achieves near-zero MSE beyond 12 dB SNR. Even 8-bit quantization maintains low MSE  at higher SNR values, resulting in further reducing the transmission data size and bandwidth saving. For QAM16, similar trends are observed with MSE values dropping to near-zero after 20 dB SNR, which is due to the increased complexity of QAM16 modulation. Overall, the Quantization approach demonstrates robust performance, especially at higher bit depths, with consistent MSE reduction as SNR improves.

Figures~\ref{fig:IEEE} displays the MSE performance for the IEEE 754 encoding scheme. For BPSK,
The MSE remains extremely high (on the order of $10^{18}$–$10^{19}$), showing that this encoding is highly sensitive to channel noise and performs poorly in low-SNR regimes. For QAM16, The MSE starts high around $10^4$ and decreases as SNR increases, reaching around zero at 22 dB. These results indicate that IEEE 754 encoding is highly susceptible to bit errors, resulting in significantly higher MSE, especially in low to moderate SNR conditions. In contrast, quantization encoding demonstrates much greater robustness and efficiency for transmitting embedding vectors, offering substantially lower MSE and improved scalability across different modulation schemes.

\begin{figure}
    \centering  
    \begin{subfigure}[b]{0.32\textwidth}
        \includegraphics[width=\linewidth]{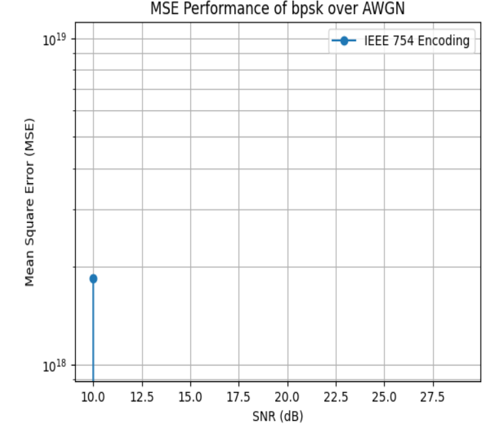}
       % \caption{Caption 3}
        \label{fig:sub3}
    \end{subfigure}
   \hfill
    \begin{subfigure}[b]{0.35\textwidth}
        \includegraphics[width=\linewidth]{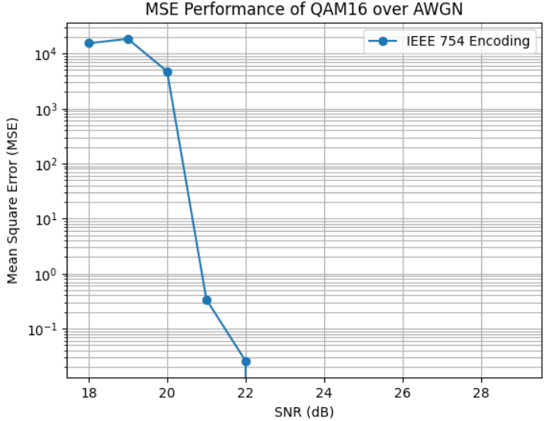}
        %\caption{Caption 4}
        \label{fig:sub4}
    \end{subfigure}
    
    \caption{Comparison of MSE in reconstructed embeddings across different modulation schemes for IEEE 754 encoding.}
    \label{fig:IEEE}
\end{figure}

Figure~\ref{fig:visual} presents a visual comparison of original and reconstructed images transmitted using two different bitstream encoding methods, 8-bit Quantization and IEEE 754 floating-point encoding, under varying SNR conditions ranging from 1 dB to 12 dB. As illustrated in Figures~\ref{fig:Quant} and~\ref{fig:IEEE}, for the BPSK modulation scheme, the MSE decreases with increasing SNR and becomes constant beyond 12~dB. This saturation indicates that further increases in SNR do not yield noticeable performance improvements. Therefore, in our experiments, we report results up to 12~dB to maintain clarity and avoid redundancy. The leftmost column shows the original image captured before transmission, serving as a reference for evaluating reconstruction quality. The middle section illustrates the performance of 8-bit quantization encoding. As SNR increases from 1 dB to 12 dB, the reconstructed image quality improves progressively. Even at low SNRs such as 3-4 dB, the structure and semantic content of the image are fairly well preserved. At higher SNRs (greater than 7dB), the images are almost indistinguishable from the original, indicating strong robustness to noise and reliable image reconstruction. The rightmost section shows the results of using IEEE 754 encoding. At low SNR values ( from 1db to 6 dB), the reconstructed images suffer from severe degradation, with high levels of pixel distortion and semantic loss. Even at moderate SNRs (from 7–9 dB), image quality remains significantly impaired. Only at higher SNRs (greater than  10dB) does the reconstruction begin to retain some meaningful visual structure. Thus, with 8-bit quantization, we can further reduce the transmission data size by 75\% compared to IEEE754 bitstream encoding.
\begin{figure}
    \centering
    \includegraphics[width=0.85\linewidth]{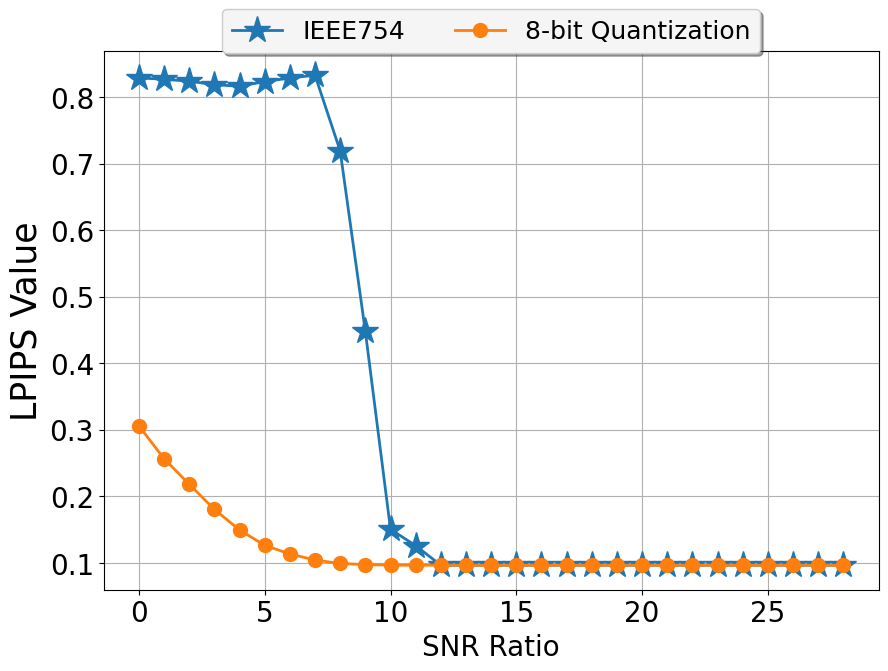}
    \caption{LPIPS value of the reconstructed images}
    \label{fig:LPIPS}
\end{figure}

\begin{figure*}
    \centering
    \includegraphics[width=0.9\linewidth]{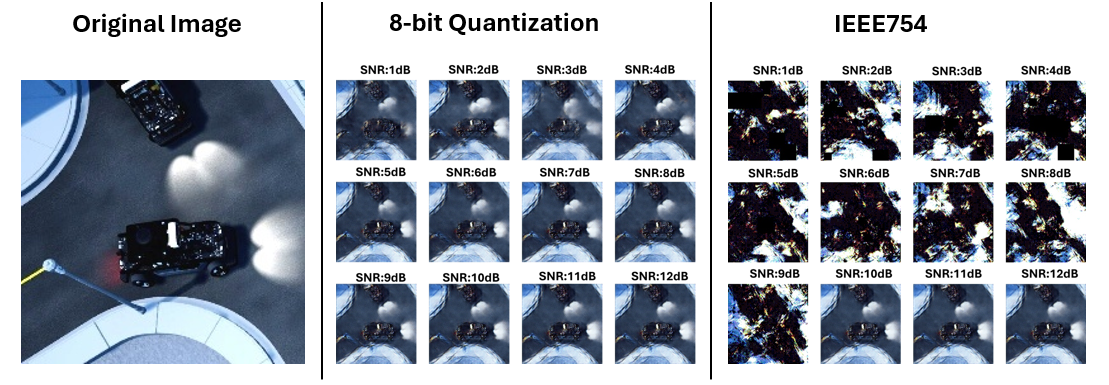}
    \caption{Visual comparison of original and reconstructed images transmitted using 8-bit Quantization and IEEE 754 bitstream encoding across varying SNR levels}
    \label{fig:visual}
\end{figure*}
Figure~\ref{fig:LPIPS} presents the perceptual similarity between the original and reconstructed images using the LPIPS metric. An LPIPS score close to 0 indicates high perceptual similarity, while a score approaching 1 reflects significant perceptual differences. It can be observed that with 8-bit quantization encoding, the LPIPS score drops to approximately 0.1 at a lower SNR of around 6 dB, compared to the IEEE 754 floating-point encoding, which achieves a similar perceptual quality only at a higher SNR of about 12 dB. These findings are consistent with the visual comparisons illustrated in Figure~\ref{fig:visual}, further confirming the efficiency of 8-bit quantization in preserving perceptual quality at lower SNRs.

\begin{figure}
    \centering
    \includegraphics[width=0.85\linewidth]{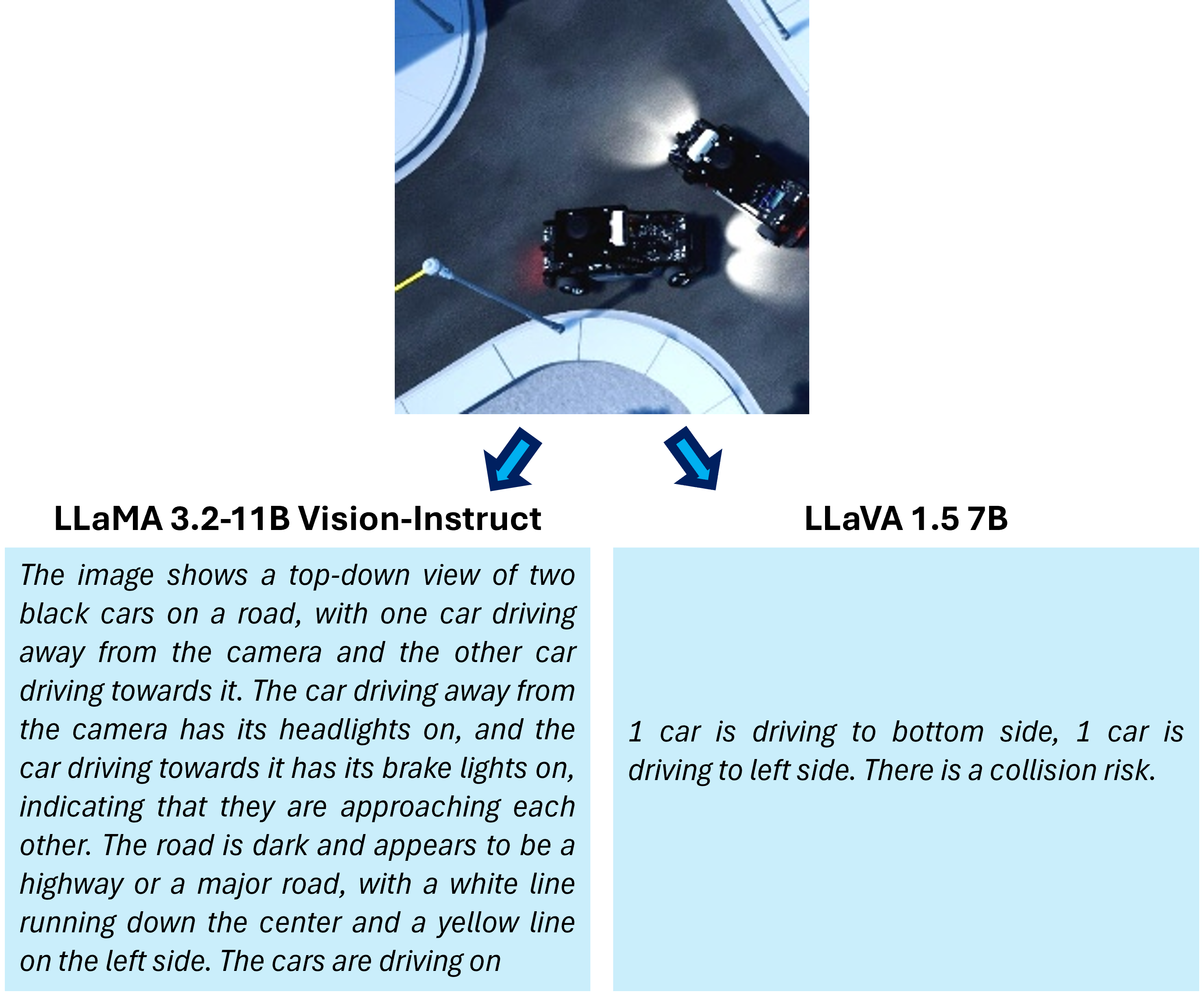}
    \caption{LLM LLaVA and LLaMA Output Comparison}
    \label{fig:example}
\end{figure}

Furthermore, the accuracy of multimodal LLM-generated responses to queries based on reconstructed images plays a crucial role in effective urban traffic monitoring. As discussed in Section~\ref{llm}, the LLaVA-1.5 7B model proves highly effective for responding to real-time traffic queries. For comparative evaluation, we also fine-tuned the LLaMA 3.2-11B Vision-Instruct model using the same dataset. The outputs of both multimodal LLMs are illustrated in Figure~\ref{fig:example}, where the left panel displays the response from the LLaMA model and the right panel shows the LLaVA output for the same query, ``Explain the vehicular activity." While both models generate contextually accurate responses, the LLaMA model tends to produce highly detailed descriptions, even when the input image is a cropped RoI. For instance, it outputs phrases such as ``The road is dark and appears to be a highway or a major road, with a white line running down the center and a yellow line on the left side," and "top-down view of two black cars on a road." Although these details are visually accurate, setting a maximum token limit can lead to truncated or incomplete responses (e.g., "The cars are driving on ..."), potentially missing the core of the query. In contrast, the LLaVA model offers concise answers such as ``1 car is driving to bottom side, 1 car is driving to left side. There is a collision risk," effectively summarizing key vehicular dynamics and potential risks. This level of brevity and relevance is particularly advantageous for real-time traffic applications, where rapid interpretation and decision-making are critical.

Quantitatively, the LLaVA model achieves 93.2\% accuracy when processing original cropped images. However, this accuracy drops slightly to 89\% when reconstructed images are used. This trade-off suggests that, despite minor performance degradation, the use of image embedding representations can still be a viable approach for semantic information transmission in memory-constrained or bandwidth-limited scenarios, offering a balance between efficiency and inference accuracy.

\section{Conclusions and Future Work} \label{sec:conclusion}

This paper has presented a semantic communication framework for urban traffic management using multimodal LLMs. The proposed approach has enabled fast and efficient communication suitable for real-time traffic monitoring applications. Key visual Regions of Interest have been identified using the YOLO model, and the corresponding image patches have been cropped and resized for input into a multimodal LLM. To optimize transmission efficiency within the semantic communication framework, ViT has been employed to convert each cropped image into embedding vector representations. These embeddings have then been encoded using both IEEE 754 floating-point and quantization-based bitstream encoding techniques for transmission. Experimental results, including MSE and LPIPS scores, have demonstrated that the quantization technique has offered superior robustness to bit errors and has further reduced transmission data size compared to IEEE 754 encoding. On the cloud server, the decoder has successfully reconstructed the image segments from their embeddings, which have subsequently been processed by the LLaVA 1.5 7B multimodal LLM to generate accurate and context-aware traffic responses.

%In this research, both the transmission of embedding vectors and the direct cropped image transmission approaches have been evaluated in terms of their impact on transmission efficiency and multimodal LLM performance. When using cropped images instead of ViT-based embedding, the system has achieved approximately 93\% accuracy with the LLaVA 1.5 model and demonstrated 93–98\% reduction in transmission data size. In contrast, converting cropped images into embedded vectors has further reduced the transmission load, achieving up to 99.9\% reduction in transmission data size. However, this has come at the cost of a slight performance drop of the LLaVA 1.5 model, with accuracy reduced to 89\%.
In this work, we evaluated both the transmission of embedding vectors and the direct transmission of cropped images, examining their effects on transmission efficiency and multimodal LLM performance. Transmitting cropped images achieved approximately 93\% accuracy with the LLaVA 1.5 model while reducing transmission data size by 93–98\%. Converting these cropped images into ViT-based embedding vectors further decreased the transmission load, achieving up to 99.9\% reduction in data size. However, this extreme compression introduced a modest accuracy drop to 89\%, highlighting the inherent trade-off that greater bandwidth savings come at the expense of a slight reduction in model performance.
Future work will explore a broader range of text-based queries, and we also plan to investigate the predictive capabilities of the proposed framework for anticipating future events based on visual inputs. Moreover, we will explore more diverse vision-grounded LLMs, such as BLIP-2 and Qwen-VL, for traffic queries to enhance the performance. Moreover, context awareness can be integrated to enhance the inference performance in real world deployment. 
%This research mainly focused on the vehicles in a fixed scale surrounding. Future extensions include introducing a dynamic scale for each vehicle depending on its state in the simulation to obtain inferences with higher accuracy on vehicle activities. %Moreover, we will include pedestrians, but using a different cropping technique due to the possibility of a vast number of pedestrians and their overlapping windows of RoI.  

\section*{Acknowledgment }\label{Section6}
This work is supported in part by the MITACS Accelerate Program, in part by the Natural Sciences and Engineering Research Council of Canada (NSERC) CREATE TRAVERSAL program, and in part by the Ontario Research Fund-Research Excellence program under grant number ORF-RE012-026. The authors would like to thank Quanser for their support in the generation of the traffic data via Quanser Interactive Lab (https://www.quanser.com/products/quanser-interactive-labs/).

%----------------------------------------------------------------------------------------
%	BIBLIOGRAPHY
%----------------------------------------------------------------------------------------

%\bibliography{biblio.bib}{}

\bibliographystyle{IEEEtran}
%\bibliography{biblio.bib}{}


\end{document}